\documentclass[a4paper,preprint,review,authoryear,1p,12pt]{elsarticle}
\usepackage{amssymb}
\usepackage{color}
\usepackage{subfigure}
\usepackage{amsmath}      
\usepackage{array}    
\usepackage{framed} 
\usepackage[titletoc]{appendix}
\usepackage{amsthm}
\usepackage[figuresright]{rotating}
\usepackage{float}
\usepackage{soul}
\usepackage{algorithmicx}
\usepackage{algpseudocode}
\biboptions{numbers,comma,square,sort&compress}
\usepackage[bookmarksnumbered,colorlinks,linkcolor=red,anchorcolor=blue,citecolor=green]{hyperref}
\usepackage{geometry}
\usepackage{fancyhdr}
\graphicspath{{figures/}}
\usepackage{multirow} 
\usepackage{booktabs} 
\usepackage{caption}
\usepackage{epstopdf}

\usepackage{graphics}
\usepackage{epsfig}
\usepackage[ruled,linesnumbered]{algorithm2e}

\newtheorem{remark}{Remark}

\begin{document}

\begin{frontmatter}

	\title{{ Dynamic fault detection and diagnosis for alkaline water electrolyzer
			with variational Bayesian Sparse principal component analysis}}

	\author[label1]{Qi Zhang }
	\author[label1]{Weihua Xu}
	\author[label1]{Lei Xie \corref{cor1}}
	\author[label1]{Hongye Su}

	\address[label1]
	{State Key Laboratory of Industrial Control Technology, Zhejiang University, 310027 Hangzhou, China}
	\cortext[cor1]{Corresponding author}

\begin{abstract}

Electrolytic hydrogen production serves as not only a vital source of green hydrogen but also a key strategy for addressing renewable energy consumption challenges. 
For the safe production of hydrogen through alkaline water electrolyzer (AWE), dependable process monitoring technology is essential. 
However, random noise can easily contaminate the AWE process data collected in industrial settings, presenting new challenges for monitoring methods. 
In this study, we develop the variational Bayesian sparse principal component analysis (VBSPCA) method for process monitoring. 
VBSPCA methods based on Gaussian prior and Laplace prior are derived to obtain the sparsity of the projection matrix, which corresponds to $\ell_2$ regularization and $\ell_1$ regularization, respectively.
The correlation of dynamic latent variables is then analyzed by sparse autoregression and fault variables are diagnosed by fault reconstruction.
The effectiveness of the method is verified by an industrial hydrogen production process, and the test results demonstrated that both Gaussian prior and Laplace prior based VBSPCA can effectively detect and diagnose critical faults in AWEs.

\end{abstract}


\end{frontmatter}

\section{Introduction}

As the global energy transition accelerates,  Alkaline water electrolyzer (AWE) for hydrogen production has become the most widely used hydrogen production technology.  
As industrial hydrogen production grows in scale, the safety of hydrogen production has attracted more attention.
The process of producing hydrogen involves multiple scientific disciplines, and the system is particularly complex and difficult to control.
With the widespread adoption of Distributed control systems (DCS) and sensor technology in industrial settings, it is possible to collect and store large amounts of data from the AWE process.
Moreover, data mining and database technologies offer strong technical support for developing and applying data-driven modeling approaches in industrial AWE processes\cite{WISE1996329}. 
Hence, by extracting and analyzing information from process data, process monitoring models can be built for industrial applications to ensure safe operation of industrial water electrolyzer processes\cite{raveendranProcessMonitoringUsing2018,MACGREGOR1995403}.

Latent variable models such as Principal component analysis (PCA) and Partial least squares (PLS) enable the extraction of key information from the process to build efficient monitoring systems by projecting high-dimensional process data into a low-dimensional space\cite{qinSurveyDatadrivenIndustrial2012a}.
However, both PCA and PLS under the assumption that the data involved in the process exhibit no temporal correlation.
 In fact, there is significant dynamic correlation in the process data. 
Therefore, neither of these methods exploits the dynamic correlation between the data. 
There are three types of Dynamic latent variable (DLV) methods that have been most extensively studied\cite{qinBridgingSystemsTheory2020a}. 
The most conventional approach is to construct time lag matrices, such as the dynamic PCA method, but this approach significantly reduces the interpretability of faults\cite{kuDisturbanceDetectionIsolation1995d,6718065}.
In addition, dynamic relationships can be modelled explicitly by state space models, such as variational Bayesian state space models\cite{zhangVariationalBayesianState2023}. 
There are also methods such as dynamic inner PCA \cite{dongNovelDynamicPCA2018a} and dynamic inner canonical correlation analysis (DiCCA) \cite{dongNewDynamicPredictive2020a} which focus more on the dynamic relationships between latent variables.
However, the disadvantage of the latent variable models is the high requirement for data quality\cite{zhangQualityRelevantProcessMonitoring2022}. 
The latent variable models look for the orthogonal direction with the largest data variance to separate noise and data as much as possible\cite{liuRobustVariationalInference2021}. 
Given that industrial process data are easily contaminated by noise, latent variable models are sensitive to noise and outliers, which reduces the reliability of monitoring.

Sparse methods have a strong noise reduction capability that removes noise from the data while retaining critical information\cite{candesIntroductionCompressiveSampling2008,natarajanSparseApproximateSolutions1995}. 
This is especially important for fault detection in noisy industrial environments.
Traditionally, sparsity is obtained by utilizing regularization in the loss function. $\ell_1$ regularization and $\ell_2$ regularization are the most common methods, corresponding to Lasso regression and Ridge regression, respectively\cite{muthukrishnanLASSOFeatureSelection2016}.
The core of Lasso regression is that it uses the sum of the absolute values of the weight coefficients as the penalty term, which makes the loss function of Lasso regression discontinuously derivable\cite{tibshiraniRegressionShrinkageSelection1996}.
The selection of the regularization factor is usually difficult, and it is empirically selected, resulting in non-robustness to changes in the algorithmic environment.
Bayesian sparsity treats regularization as a Bayesian framework for prior knowledge\cite{jiBayesianCompressiveSensing2008,guanSparseProbabilisticPrincipal}.
In this approach, the parameter estimates are updated with the posterior distributions, treating the regularisation terms as prior distributions of the model parameters\cite{babacanBayesianCompressiveSensing2010}.
In addition, sparse Bayesian learning (SBL) has become a hot topic of research because it improves the interpretability and generalisability of the model by taking parameter uncertainty into account\cite{tipping2001sparse,wipfSparseBayesianLearning2004}.
Expectation maximisation (EM), which reduces computational cost and handles missing values, is a common approach to implementing sparse Bayesian methods\cite{bishopPatternRecognitionMachine,sammaknejadReviewExpectationMaximization2019}.
However, EM can only provide point estimates of parameters that may fall into local optima or overfitting, and requires strict assumptions about the posterior distributions of the hidden variables. Therefore, the full Bayesian uncertainty of the model cannot be captured by the EM method\cite{bleiVariationalInferenceReview2017,jordanIntroductionVariationalMethods1999b}.
A more advantageous approach involves utilizing variational Bayesian methods, which approximate the complete posterior distribution of parameters and hidden variables, thereby circumventing the constraints associated with the EM algorithm\cite{bealVariationalBayesianLearning2006,bealVariationalAlgorithmsApproximate2003}.

The main objective of this study is to develop a process monitoring technique for the AWE hydrogen production process that can effectively handle noise.
To achieve this, we have derived variational Bayesian principal component analysis (VBSPCA) methods for process monitoring based on Gaussian and Laplace priors, respectively, inspired by Bayesian sparsity.
The introduction of $\ell_2$ regularization is equivalent to introducing a Gaussian prior, while the introduction of $\ell_1$ regularization is equivalent to introducing a Laplace prior.
By employing matrices with a shared sparse structure and independent Gaussian priors for low-rank decomposition, Gaussian prior-based VBSPCA is capable of accurately modeling and estimating noise, thereby mitigating its effects. 
The Laplace prior-based VBSPCA designs a two-level hierarchical decomposition of the Laplace distribution to construct the conjugate structure for Bayesian posterior inference.
In addition, the parameters of VBSPCA are estimated by variational Bayesian inference, which not only quantifies the model uncertainty, but also makes the algorithm easy to handle.
Finally, to make full use of the structurally relevant information of the variables, we incorporate the vector autoregression (VAR) with $\ell_1$ regularization to better extract dynamic features from the latent variables.
The latent variables processed by VBSPCA can reduce the influence of noise. Dynamic analysis of these latent variables can improve the reliability of fault detection and diagnosis.
The effectiveness of the proposed method is verified by an industrial hydrogen production process, and the test results show that both Gaussian prior and Laplace prior-based VBSPCA can effectively detect and diagnose critical faults in AWEs.

The main contributions of this work are summarized as follows:
\begin{itemize}
	\setlength{\itemsep}{0pt}
	\setlength{\parsep}{0pt}
	\setlength{\parskip}{0pt}
	\item [1)] 
	This study develops VBSPCA techniques for AWE process diagnosis and fault detection, based on Gaussian prior and Laplace prior, respectively.
	\item [2)] 
	The developed method constructs a sparse prior that naturally reduces the impact of noise and quantifies uncertainty through variational inference.
	\item [3)] 
    $\ell_1$ regularized vector autoregression extracts dynamic features from latent variables, which can exploit structural information and reduce noise disturbances.
\end{itemize}

The remainder of the paper is organized as follows. The related works are described in Section~\ref{sec:related}. The variational Bayesian Sparse PCA method is described in Section~\ref{sec:algorithms}. The variational inference for these two VBSPCA techniques is presented in Section~\ref{sec:VBI},
while dynamic fault detection and diagnosis approach are described in Section~\ref{sec:fault detection}.
Simulation results are provided in Section~\ref{sec:experiments}, followed by concluding remarks in Section~\ref{sec:conclusions}.

\section{Related work}\label{sec:related}
\subsection{PCA and Probabilistic PCA}

Let $\boldsymbol{x}$ denote a sample vector of $m$ sensors. Assuming that there are $n$ samples for each sensor, a data matrix $\boldsymbol{X}=[\boldsymbol{x}_1,\boldsymbol{x}_2,\dots,\boldsymbol{x}_n ] \in \mathbb{R}^{m \times n} $ is composed with each column representing a sample $\boldsymbol{x}_{i}\in \mathbb{R}^{m}$.
With the projection matrix $\boldsymbol{P}\in \mathbb{R}^{m \times r},(r<m)$ satisfies $\boldsymbol{P}^{\top}\boldsymbol{P}=\boldsymbol{I}_r$, the PCA model can be obtained
\begin{align} \label{1}
	\boldsymbol{x}_i= \boldsymbol{P}\boldsymbol{t}_i +\boldsymbol{\bar{x}}+\boldsymbol{\xi}_i ,\ \ \  i = 1,2,\dots,n 
\end{align}
where $\boldsymbol{t}_i$ denotes the latent variable with $r$ dimensions,  $ \boldsymbol{\bar{x}}=\frac{1}{n}\sum\nolimits_{i=1}^{n} \boldsymbol{x}_{i} \in \mathbb{R}^{m}$ is the mean vector,
$\boldsymbol{\xi}_i\sim \mathcal{N}(\boldsymbol{0},{\boldsymbol{\sigma} }^{2}\boldsymbol{I})$ represents additive process noise.

The PCA model can be reformulated by creating matrices $\boldsymbol{T}=\left(\boldsymbol{t}_1,\boldsymbol{t}_2,...,\boldsymbol{t}_n \right) \in \mathbb{R}^{r \times n} $, $\boldsymbol{\bar{X}}=\left(\boldsymbol{\bar{x}}_1,\boldsymbol{\bar{x}}_2,...,\boldsymbol{\bar{x}}_n \right) \in \mathbb{R}^{m \times n}$, and $\Xi=\left(\boldsymbol{\xi}_1,\boldsymbol{\xi}_2,...,\boldsymbol{\xi}_n \right) \in \mathbb{R}^{m \times n}$, PCA can be rewritten as the following
\begin{align} \label{2}
	\boldsymbol{X}= \boldsymbol{P}\boldsymbol{T} +\boldsymbol{\bar{X}}+\boldsymbol{\Xi}.
\end{align}

In order to maximize the variance of the projected data,  PCA can obtain the optimal $\boldsymbol{P}$ and $\boldsymbol{T}$ by optimizing the following objective function
\begin{align}  \label{3}
	\underbrace{\arg \min}_{\boldsymbol{P},\boldsymbol{T}} | | \boldsymbol{X} - \boldsymbol{\bar{X}} - \boldsymbol{P}\boldsymbol{T} | | _ { F } ^ { 2 } \quad  \operatorname{s.t.} \boldsymbol{P} ^ { \top } \boldsymbol{P} = \boldsymbol{I}_r 
\end{align}
where $||\cdot||_F$ is the Frobenious norm. 

\begin{remark}
	In general, the optimal $\boldsymbol{P}$ can be obtained by stacking the $r$ singular vectors corresponding to the first $r$ largest singular values of the sample covariance matrix $\boldsymbol{S}=\sum\nolimits_{i=1}^{n}(\boldsymbol{x}_i-\boldsymbol{\bar{x}})(\boldsymbol{x}_i-\boldsymbol{\bar{x}})^{\top}/(n-1)$. As a result, latent variable $\boldsymbol{t}_i=(\boldsymbol{x}_i-\boldsymbol{\bar{x}})\boldsymbol{P}$ can be obtained\cite{joeqinStatisticalProcessMonitoring2003a}.
\end{remark}

The disadvantage of PCA is that it is not possible to obtain probability distributions with respect to the projection matrix $\boldsymbol{P}$ and the latent variable $\boldsymbol{t}$. 
 Probabilistic PCA (PPCA) can be obtained from a Gaussian latent variable model, and its parameters are estimated iteratively using the EM method\cite{tippingProbabilisticPrincipalComponent1999,choiFaultDetectionBased2005}.
The conditional probability distribution of $ \boldsymbol{x}_i $ can be expressed as
\begin{align}  \label{4}
	p(\boldsymbol{x}_i|\boldsymbol{t}_i)=\mathcal{N}(\boldsymbol{x}_i|\boldsymbol{P}\boldsymbol{t}_i+\boldsymbol{\bar{x} },{{\boldsymbol{\sigma }}^{2}}\boldsymbol{I}).
\end{align}

The prior distribution of the latent variable $\boldsymbol{t}_i$ is assumed to be a standard Gaussian distribution, the marginal distribution of $ \boldsymbol{x}_i $ can thus be calculated by integrating out the latent variables
\begin{align}
	p(\boldsymbol{x}_i)=\int{p(\boldsymbol{x}_i|\boldsymbol{t}_i)p(\boldsymbol{t}_i)d\boldsymbol{t}_i}=\mathcal{N}(\boldsymbol{x}_i|\boldsymbol{\bar{x}} ,\boldsymbol{P}{\boldsymbol{P}^{\top}}+{\boldsymbol{\sigma }^{2}}\boldsymbol{I}).
\end{align}

According to the Maximum likelihood estimation (MLE), the log-likelihood function is given by
\begin{align}
	\mathcal{L} ( \boldsymbol{P} , \boldsymbol{\bar{x}} , \boldsymbol{\sigma} ^ { 2 } \boldsymbol{I}| \boldsymbol{X} ) = &\sum _ { i = 1 } ^ { n } \ln p ( \boldsymbol{x} _ { i } ) 
	= \sum _ { i = 1 } ^ { n } \ln \mathcal{N} \left( \boldsymbol{x}_i|\boldsymbol{\bar{x}} ,\boldsymbol{P}{\boldsymbol{P}^{\top}}+{\boldsymbol{\sigma }^{2}}\boldsymbol{I}\right) .
\end{align}

Maximizing $\mathcal{L} ( \boldsymbol{P} , \boldsymbol{\bar{x}} , \boldsymbol{\sigma} ^ { 2 } \boldsymbol{I}| \boldsymbol{X} )$ yields the optimal $\boldsymbol{P}$, $\boldsymbol{\bar{x}}$, and ${\boldsymbol{\sigma }^{2}}$.

\begin{remark}
	Finding the projection matrix $\boldsymbol{P}$, where $r<m$ accomplishes the reduced dimensionality and feature extraction, is the goal of both PCA and PPCA. 
	In this way, the variable $\boldsymbol{x}_i$ can be mapped to the low-dimensional space and the latent variable (extracted feature) can be obtained.
	This is actually the process of feature extraction.
\end{remark}

	Typically, the matrix $\boldsymbol{P}$ derived from PCA and PPCA is not sparse, leading to poor interpretability. If $\boldsymbol{P}$ is sparse, then the latent variable $\boldsymbol{t}$ can be expressed as a weighted combination of a small number of original features, thereby improving the interpretability of the model.

\subsection{Sparsity analysis}
To obtain sparse solutions for $\boldsymbol{P}$, Lasso regression ($\ell_1$ regularization) and Ridge regression ($\ell_2$ regularization) are commonly used. 
The optimization objective of Lasso regression is expressed as
\begin{align} \label{7}
	\boldsymbol{P}^{\ast} = \underbrace{\arg \min}_{\boldsymbol{P} } \frac{1}{n} \sum\limits_{i=1}^{n} ||  \boldsymbol{x}_i - \boldsymbol{P}\boldsymbol{t}_i   ||+ \lambda ||\boldsymbol{P} ||_1 .
\end{align}

The optimization objective of Ridge regression is expressed as
\begin{align}  \label{8}
	\boldsymbol{P}^{\ast} = \underbrace{\arg \min}_{\boldsymbol{P} } \frac{1}{n} \sum\limits_{i=1}^{n} ||  \boldsymbol{x}_i -\boldsymbol{P}\boldsymbol{t}_i   ||+ \lambda ||\boldsymbol{P} ||_2^2 .
\end{align}

\begin{remark}
	Using the $\ell_1$ norm as a regularization term allows for feature selection, resulting in a sparse model where some regression coefficients can be compressed to zero; while using the $\ell_2$ norm as a regularization term can help prevent overfitting.
\end{remark}

Following the conditional probability distribution of $ \boldsymbol{x}_i $ in \eqref{4}, sparsity analysis will be reformulated as maximizing the posterior probability
\begin{align}  \label{9}
	\begin{split}
		\boldsymbol{P}^{\ast} &= \underbrace{\arg \max}_{\boldsymbol{P} }  \ln \left( \prod _ { i = 1 } ^ { n }  p(\boldsymbol{x}_i|\boldsymbol{t}_i, \boldsymbol{P} ) p(\boldsymbol{P})   \right) \\
		&= \underbrace{\arg \max}_{\boldsymbol{P} } \sum\limits_{i=1}^{n} \left(    \ln p(\boldsymbol{x}_i|\boldsymbol{t}_i, \boldsymbol{P} )+ \ln    \sum\limits_{i=1}^{n} p(\boldsymbol{p}_i)               \right) 
	\end{split}
\end{align}
where $ \ln p(\boldsymbol{x}_i|\boldsymbol{t}_i, \boldsymbol{P} ) \propto \left( \boldsymbol{x}_i - \boldsymbol{P}\boldsymbol{t}_i \right)^2 $. 

It can be seen that the target loss part of \eqref{7} is solved here by a probability  distribution.
The significance of $p(\boldsymbol{P})$ is the prior assumption about the probability distribution of $\boldsymbol{P}$. The Ridge Regression will be obtained by assuming that the prior distribution of $\boldsymbol{P}$ is a zero-mean Gaussian distribution $P ( \boldsymbol{p} _ { i } ) = \frac { {\gamma} _ { i } } { \sqrt { 2\pi } } \exp ( - {\gamma} _ { i } | |  \boldsymbol{p} _ { i } || ^ { 2 } )$.
Correspondingly, \eqref{9} can be calculated as
\begin{align}
	\boldsymbol{P}^{\ast} =\underbrace{\arg \max}_{\boldsymbol{P} } \sum\limits_{i=1}^{n} \left(    \ln p(\boldsymbol{x}_i|\boldsymbol{t}_i, \boldsymbol{P} )+   \boldsymbol{\gamma} \sum\limits_{i=1}^{n} ||\boldsymbol{p}_i ||^2      \right) .
\end{align}

The Lasso Regression will be obtained by assuming that the prior distribution of $\boldsymbol{P}$ is a zero-mean Laplace distribution $P ( \boldsymbol{p} _ { i } ) = \frac { \varphi_i} { 2 } \exp ( - \varphi | \boldsymbol{p}_i | )$,
Correspondingly, \eqref{9} can be calculated as
\begin{align}
	\boldsymbol{P}^{\ast} =\underbrace{\arg \max}_{\boldsymbol{P} } \sum\limits_{i=1}^{n} \left(    \ln p(\boldsymbol{x}_i|\boldsymbol{t}_i, \boldsymbol{P} )+   \boldsymbol{\varphi} \sum\limits_{i=1}^{n} |\boldsymbol{p}_i |     \right) .
\end{align}

Consequently, it can be shown that in Bayesian inference, the Laplace prior is in fact equivalent to $\ell_1$ regularization, and the Gaussian prior is equivalent to $\ell_2$ regularization. It can usually be solved by EM methods, however, as a point estimation method, the EM method can only estimate the maximum posterior of the single most probable value for each parameter. This parameter learning method may lead to local optima and overfitting, thus limiting the performance of the model. 

\section{Variational Bayesian Sparse PCA} \label{sec:algorithms}

\subsection{Gaussian prior-based VBSPCA} \label{sec:Model}

Suppose  $\boldsymbol{X} $ can be decomposed into a low-rank matrix $\boldsymbol{H}$ and the noise $\boldsymbol{\Xi} $
\begin{align} \label{12}
	\begin{split}
		\boldsymbol{X} &= \boldsymbol{H} + \boldsymbol{\xi}
		=\sum\nolimits_{i=1}^{k}{{{{\boldsymbol{p}}_{i}}\boldsymbol{t}_{i}^{\top }}}+ \boldsymbol{\Xi}
	\end{split}
\end{align}
where $\boldsymbol{H}=\boldsymbol{P}\boldsymbol{T}^{\top}$ such that $\mathsf{Rank}(\boldsymbol{H}) = r \ll min (n,m)$.

According to the principle of $\ell_2$ regularized sparsity,  a sparsity inducing Gaussian priors is used on each coefficient of the unknown vector based on Automatic relevance determination (ARD).
The columns of $\boldsymbol{P}$ and $\boldsymbol{T}$ are associated with Gaussian priors of precisions $\boldsymbol{\gamma}$ to ensure that the majority of the columns in each are set to zero, so $\boldsymbol{\gamma}$ is closely related to the sparsity of the solution\cite{wipfSparseBayesianLearning2004}
\begin{align}  \label{13}
	\begin{split}
		& p ( \boldsymbol{P} | \boldsymbol{\gamma} ) = \prod _ { i = 1 } ^ { k } \mathcal{N} ( \boldsymbol{p}_i | 0 , {\gamma} _ { i } ^ { - 1 } \boldsymbol{I} _ { m } ), \\ 
		& p ( \boldsymbol{T} | \boldsymbol{\gamma} ) = \prod _ { i = 1 } ^ { k } \mathcal{N} ( \boldsymbol{t}_i | 0 , {\gamma} _ { i } ^ { - 1 } \boldsymbol{I} _ { n } )
	\end{split}
\end{align}
where ${{\boldsymbol{I}}_{m}}$ is the $m \times m$ identity matrix, ${{\boldsymbol{I}}_{n}}$ is the $n \times n$ identity matrix. Thus, the columns of $\boldsymbol{P}$ and $\boldsymbol{T}$ have the same sparsity profile enforced by the common precisions ${\gamma}_i$.

In this inference process, many of the precisions $\gamma_{ i }$  will assume very large values, which effectively removes the corresponding outer-products from $\boldsymbol{H}$  and thus reduces the rank of the estimate.
Gamma priors  $p ( \gamma _ { i } )  $ can be placed on the precision parameters of the above Gaussian distributions of $\boldsymbol{P}$ and $\boldsymbol{T}$. The form of conjugate Gamma hyperprior is expressed as
\begin{align}  \label{14}
	p ( \gamma _ { i } ) = \text{Gam} ( a_0 , \frac { 1 } { b_0 } ) \propto \gamma _ { i } ^ { a_0 - 1 }  \exp ( - b_0 \gamma _ { i } )
\end{align}

\begin{remark}
	To ensure that the hyperpriors are sufficiently vague and diffuse, we treat the parameters $a_0$ and $b_0$ as fixed constants and assign them very small values, such as $10^{-5}$, instead of estimating them from the data.
\end{remark}

\begin{remark}
	The noise in the measurements is assumed to be i.i.d. drawn from a Gaussian distribution, which has an effect on each of the measurements. Noting that this construction is only within the prior, and that an approximation to the posterior $\boldsymbol{\xi}$ can be inferred for perturbations to the prior.
\end{remark}

Mathematically, the noise in \eqref{2} is modeled as
\begin{align}  \label{15}
	p(\boldsymbol{\Xi}|\beta)=\prod\limits_{j=1}^{m}{\prod\limits_{i=1}^{n}{\mathcal{N}({\boldsymbol{\Xi}_{ji}}|0,{\beta}^{-1})}}.
\end{align}

The parameter selection of the prior $\alpha$ can largely affect the parameter estimation of the posterior distribution.
 To reduce the effect of the parameter selection of the original distribution, the noise precision $\beta$ is set to the non-informative Jeffrey’s prior
\begin{align} \label{16}
	p ( \beta ) = \beta^{-1} .
\end{align}

In accordance with the form of PCA in \eqref{1}, the mean $\boldsymbol{\bar{X}}$ in \eqref{12} is modeled as
\begin{align}
	p(\boldsymbol{\bar{X}}|\boldsymbol{\alpha})=\prod\limits_{j=1}^{m}{\prod\limits_{i=1}^{n}{\mathcal{N}({\boldsymbol{\bar{X}}_{ji}}|0,{\alpha_{ji}}^{-1})}}.
\end{align}

\begin{remark}
	In classical PCA fault detection method, a preliminary step includes zero-centering. Hence, the decomposition of $\boldsymbol{X}$ doesn't consider the mean values.
	For the rigor of the algorithm, we give here the form of the mean $\boldsymbol{\bar{X}}$, which can also be interpreted as the sparse error.
\end{remark}

Similarly, the precision $\alpha$ is set to the non-informative Jeffrey’s prior
\begin{align}  \label{18}
	p ( \alpha_{ji} ) = \alpha_{ji}^{-1} .
\end{align}

The conditional probability distribution for $\boldsymbol{X}$ can be expressed as follows
\begin{align}
	p(\boldsymbol{X}|\boldsymbol{P},\boldsymbol{T},\beta)=\prod\limits_{j=1}^{m}{\prod\limits_{i=1}^{n}{\mathcal{N}({{y}_{ji}}|x_{ji},{\beta}^{-1})}} .
\end{align}

Finally, with the conditional probability and priors, the joint distribution is expressed as
\begin{align} \label{20}
	\begin{split}
		p ( \boldsymbol{X} , \boldsymbol{P} , \boldsymbol{T} ,\boldsymbol{\bar{X}} , \boldsymbol{\gamma} , \beta ) = &p ( \boldsymbol{X} | \boldsymbol{P} , \boldsymbol{T} ,\boldsymbol{\bar{X}}, \beta ) p ( \boldsymbol{P} | \boldsymbol{\gamma} ) p ( \boldsymbol{T} | \boldsymbol{\gamma} ) p(\boldsymbol{\bar{X}}|\boldsymbol{\alpha}) p(\boldsymbol{\alpha}) p (  \boldsymbol{\gamma} )  p ( \beta ) .
	\end{split}
\end{align}


\subsection{Laplace prior-based VBSPCA}

The Laplace distribution has coarser tails than the Gaussian distribution, hence the name heavy-tailed distribution. It means that the Laplace distribution can fit heavy noise and outliers better than the Gaussian distribution.

\begin{remark}
	It is necessary to employ another prior, a two-level hierarchical decomposition of the Laplace distribution, as the Laplace distribution is not a conjugate prior to the Gaussian distribution and does not support posterior inference within a Bayesian framework.
\end{remark}

According to the principle of $\ell_2$ regularized sparsity, the first level of $\boldsymbol{P}$ introduces a Gaussian prior $p(\boldsymbol{P}_{ji}| \boldsymbol{\eta} _ { j i }) \sim \mathcal{N}(0 , \boldsymbol{\eta}_{ji})$, while the second level of $\boldsymbol{P}$ is  an exponential distribution hyperprior to the variance $\boldsymbol{\eta} _ { j i } $, $p ( \boldsymbol{\eta} _ { j i } ) = \frac { 1 } { \varphi } \exp ( - \frac { \boldsymbol{\eta} _ { j i }  } { \varphi } )$
\begin{align}
	\begin{split}
		p(\boldsymbol{P}_{ji}|\boldsymbol{\eta} _ { j i }) &= \int p ( \boldsymbol{P} _ { j i } | \boldsymbol{\eta} _ { j i } ) p ( \boldsymbol{\eta} _ { j i } ) d \boldsymbol{\eta} _ { j i } \\
		&= \frac { 1 } { 2 } \sqrt { \frac { 2 } { \varphi } } \exp ( - \sqrt { \frac { 2 } { \varphi } } | \boldsymbol{P} _ { j i } | )
	\end{split}
\end{align}

Assuming that the noise $\boldsymbol{\xi}$  follows a Gaussian distribution $\boldsymbol{\xi} \sim \mathcal{N}(0,\boldsymbol{\vartheta}^{-1})$, a prior over the inverse noise variance $\boldsymbol{\vartheta}$ is introduced
\begin{align}  
	p(\boldsymbol{\vartheta})=\text{Gam}(\boldsymbol{\vartheta}|c_0,d_0)  .
\end{align}

In accordance with the form of PCA in \eqref{1}, it is assumed that both the mean $\boldsymbol{\bar{x}}$ and the latent variable $\boldsymbol{t}$ have Gaussian distributions
\begin{align}
	\begin{split}
		p(\boldsymbol{\bar{x}}) &\sim \mathcal{N}(0, \varsigma^{-1}\boldsymbol{I} )\\
		p(\boldsymbol{t}_i) &\sim \mathcal{N}(0,\boldsymbol{I} )
	\end{split}
\end{align}

Finally, with the conditional probability and priors, the joint distribution is expressed as
\begin{align} \label{24}
	\begin{split}
		p ( \boldsymbol{X} , \boldsymbol{P},\boldsymbol{t}, \boldsymbol{\bar{x}}, \boldsymbol{\vartheta} ) = 
		\prod\limits_{i=1}^{n} p( \boldsymbol{y} | \boldsymbol{P},\boldsymbol{t}, \boldsymbol{\bar{x}}, \boldsymbol{\vartheta}, \boldsymbol{\eta}
		)P(\boldsymbol{t})p(\boldsymbol{P}|\boldsymbol{\eta})p(\boldsymbol{\eta})p(\boldsymbol{\bar{x}})p(\boldsymbol{\vartheta})
	\end{split}
\end{align}

\section{Variational Bayesian inference}\label{sec:VBI}
\subsection{A review of variational inference}
A variational Bayesian approach is implemented in this section to approximate the posterior distribution of the parameters.
The variational Bayesian of learning $	p(\boldsymbol{\Theta}|\boldsymbol{y})$ avoids intractable integration by learning an approximating distribution $Q(\boldsymbol{\Theta})$.
As a result, the marginal log-likelihood function $\ln p(\boldsymbol{y}$) can be separated into two components: the Evidence Lower Bound (ELBO) $\mathcal{L}(Q(\boldsymbol{\Theta}))$ and Kullback-Leibler (KL) divergence $\mathbb{KL}(Q(\boldsymbol{\Theta})||p(\boldsymbol{\Theta}|\boldsymbol{y}))$\cite{bealVariationalAlgorithmsApproximate2003}
\begin{align} \label{11}
	\begin{split}
		\ln p(\boldsymbol{y})&=\mathcal{L}(Q(\boldsymbol{\Theta}))+\mathbb{KL}\left( Q(\boldsymbol{\Theta})||p(\boldsymbol{\Theta}|\boldsymbol{y})\right)  \\
		&= \ln \int_{\boldsymbol{\Theta}} Q(\boldsymbol{\Theta})  \left\lbrace \frac{p(\boldsymbol{\Theta},\boldsymbol{y})}{Q(\boldsymbol{\Theta})} \right\rbrace d \boldsymbol{\Theta} - \int_{\boldsymbol{\Theta}}Q(\boldsymbol{\Theta}) \ln \left\lbrace \frac{p(\boldsymbol{\Theta},\boldsymbol{y})}{Q(\boldsymbol{\Theta})} \right\rbrace d \boldsymbol{\Theta}
	\end{split}
\end{align}
where $Q(\boldsymbol{\Theta})$ approximates as closely as possible the true distribution $p(\boldsymbol{\Theta}|\boldsymbol{y})$ through KL divergence.

\begin{remark}
	 Since $KL(Q\parallel p)\ge 0$, $ \mathcal{L}(Q(\boldsymbol{\Theta})) $ is a rigorous lower bound for $ \ln p\left( \boldsymbol{y} \right) $.
	Maximizing $ \mathcal{L}(Q(\boldsymbol{\Theta})) $ is equivalent to minimizing $ KL\left( Q\parallel p \right) $, and thus the posterior distribution $ p(\boldsymbol{\Theta} |\boldsymbol{y}) $ can be approximated by $ {Q}\left( \boldsymbol{\Theta}  \right) $.
\end{remark}

With the Mean field variational inference (MFVI), the free joint distribution is chosen to be factorable \cite{opperAdvancedMeanField2001}
\begin{align} \label{23}
		Q(\boldsymbol{\Theta})= \prod \limits_{\text{i}=1}^{I}q_i(\Theta_i). 
\end{align}

In order to find an approximate posterior distribution for each latent variable, the expected values of all parameters in the joint distribution are relative to their most recent distribution.
The posterior approximation $q_i(\Theta_i)$ for each latent variable $\Theta_i$ is optimized by
\begin{align}
	q_j(\Theta_j)=\frac{ \exp \mathbb{E}_{i\neq j}(\ln p(\boldsymbol{y},\boldsymbol{\Theta})) }{\int_{\Theta_j} \exp \left(  \mathbb{E}_{i\neq j}(\ln p(\boldsymbol{y},\boldsymbol{\Theta}))  \right) } .
\end{align}
where $\mathbb{E}(\cdot)$ is the expectation.

\subsection{Variational inference for Gaussian prior-based VBSPCA}

The latent variable and parameters of the Gaussian prior-based VBSPCA are $\boldsymbol{\Theta} = \left(  \boldsymbol{P},\boldsymbol{T}, \boldsymbol{\bar{X}}, \boldsymbol{\gamma}, \boldsymbol{\alpha}, \beta \right) $. Furthermore, following \eqref{23}, $\boldsymbol{\Theta}$ can be written as 
\begin{align}
	Q(\boldsymbol{\Theta}) =q( \boldsymbol{P}) q(\boldsymbol{T}) q(\boldsymbol{\bar{X}}) q(\boldsymbol{\gamma})    q(\boldsymbol{\alpha})    q(\beta).
\end{align}

Optimizing the approximate distribution $Q(\boldsymbol{\Theta})$  based on the joint distribution of \eqref{20} gives the form of these parameters and latent variables
\begin{subequations}
	\begin{align}
		\label{a} q( \boldsymbol{p}_i) &= \mathcal{N} ( \boldsymbol{p} _ { i  } , | \boldsymbol{\mu}^{\boldsymbol{p}} _ { i  } , \boldsymbol{\Sigma}^{\boldsymbol{p}}_ { i  } )  \\
		\label{b} q( \boldsymbol{t}_j) &= \mathcal{N} ( \boldsymbol{t} _ { j  } , | \boldsymbol{\mu}^{\boldsymbol{t}} _ { j  } , \boldsymbol{\Sigma}^{\boldsymbol{t}}_ { j  } )  \\
		\label{c} q( \boldsymbol{\bar{X}}_{ji}) &=\mathcal{N} ( \boldsymbol{\bar{X}}_{j i} , | \boldsymbol{\mu}^{\boldsymbol{\bar{X}}}_{j i}  , \boldsymbol{\Sigma}^{\boldsymbol{\bar{X}}}_{j i} )  \\
		\label{d} q(\boldsymbol{\gamma})  &= \text{Gam}({\gamma}_i| a,b) 
	\end{align}
\end{subequations}

Following \eqref{16} and \eqref{18}, $\boldsymbol{\alpha}$ and $\beta$ introduce a non-informative Jeffrey’s prior with the following expectations 
\begin{align}
	\begin{split}
		\mathbb{E}(\boldsymbol{\alpha}_{ji}) &= \frac { \mathbb{E}(\beta)+\mathbb{E}(\alpha_{ji}) } { 1+ \mathbb{E}(\boldsymbol{\bar{X}}_{ji}) (\mathbb{E}(\beta)+\mathbb{E}(\alpha_{ji}))  } \\
		\mathbb{E}(\beta) &= \frac { m n } { \mathbb{E} \left(  | | \boldsymbol{X} - \boldsymbol{P} \boldsymbol{T} ^ { \top } -\boldsymbol{\bar{X}} | | _ { F } ^ { 2 } \right) }
	\end{split}
\end{align}

The approximate distribution $q( \boldsymbol{p}_i)$ in \eqref{a} is given by
\begin{align}
	\begin{split}
		{ \boldsymbol{\mu}^{\boldsymbol{p}} _ { i  }}  & =   \mathbb{E}(\beta) \boldsymbol{\Sigma}^{\boldsymbol{p}}_ { i  } \mathbb{E}({\mathbb{\boldsymbol{T}}})(\boldsymbol{y}_i-\mathbb{E}(\boldsymbol{\bar{x}}_i))
		\\
		\boldsymbol{\Sigma}^{\boldsymbol{p}}_ { i  } & = \left(  \mathbb{E}(\beta)  \mathbb{E} (\boldsymbol{T} ^ { \top } \boldsymbol{T})  + \mathbb{E}(\gamma)\boldsymbol{I} \right)  ^ { - 1 }  .
	\end{split}
\end{align}

The approximate distribution $q( \boldsymbol{t}_i)$ in \eqref{b} is given by
\begin{align}
	\begin{split}
		{ \boldsymbol{\mu}^{\boldsymbol{t}} _ { j  }}  & =   \mathbb{E}(\beta) \boldsymbol{\Sigma}^{\boldsymbol{t}}_ { j  } \mathbb{E}({\mathbb{\boldsymbol{P}}}) (\boldsymbol{y}_j-\mathbb{E}(\boldsymbol{\bar{x}}_j))
		\\
		\boldsymbol{\Sigma}^{\boldsymbol{t}}_ { j  } & = \left(  \mathbb{E}(\beta)  \mathbb{E} (\boldsymbol{P} ^ { \top } \boldsymbol{P})  + \mathbb{E}(\gamma)\boldsymbol{I} \right)  ^ { - 1 }  .
	\end{split}
\end{align}

The approximate distribution $q( \boldsymbol{\bar{X}_{ji}})$ in \eqref{c} is given by
\begin{align}
	\begin{split}
		\boldsymbol{\mu}^{\boldsymbol{\bar{X}}}_{j i}  & =  \frac{\mathbb{E}(\beta)}{\mathbb{E}(\beta)+\mathbb{E}(\alpha_{ji})} (    \boldsymbol{X}_{ji} -\mathbb{E}(\boldsymbol{p}_i )  \mathbb{E}(\boldsymbol{t}_i )^{\top}  )
		\\
		\boldsymbol{\Sigma}^{\boldsymbol{\bar{X}}}_{j i} & =\frac{1}{\mathbb{E}(\beta)+\mathbb{E}(\alpha_{ji})} .
	\end{split}
\end{align}

The approximate distribution $q(\boldsymbol{\gamma})$ in \eqref{d} is given by
\begin{align}
	\begin{split}
		a  & =     a_0 +\frac{m + n }{2} 
		\\
		b & =   b_0 + \frac{\mathbb{E}( \boldsymbol{p}_i^{\top} \boldsymbol{p}_i ) + \mathbb{E}( \boldsymbol{t}_i^{\top} \boldsymbol{t}_i )}{2}  .
	\end{split}
\end{align}

The required moments are easily evaluated using basic properties of probability distributions
\begin{align}
	\begin{split}
		\mathbb{E} (\boldsymbol{P} ^ { \top } \boldsymbol{P}) &= \mathbb{E} (\boldsymbol{P})^{\top}\mathbb{E} (\boldsymbol{P})+m\boldsymbol{\Sigma}^{\boldsymbol{p}}_ { i  }, \\
		\mathbb{E} (\boldsymbol{T} ^ { \top } \boldsymbol{T}) &= \mathbb{E} (\boldsymbol{T})^{\top}\mathbb{E} (\boldsymbol{T})+n\boldsymbol{\Sigma}^{\boldsymbol{t}}_ { i  }, \\
		\mathbb{E}( \boldsymbol{p}_i^{\top} \boldsymbol{p}_i ) &= \mathbb{E} (\boldsymbol{p}_i)^{\top}\mathbb{E} (\boldsymbol{p}_i)+m(\boldsymbol{\Sigma}^{\boldsymbol{p}}_ { i  })_{ii},\\
		\mathbb{E}( \boldsymbol{t}_i^{\top} \boldsymbol{t}_i ) &= \mathbb{E} (\boldsymbol{t}_i)^{\top}\mathbb{E} (\boldsymbol{t}_i)+n(\boldsymbol{\Sigma}^{\boldsymbol{t}}_ { i  })_{ii},\\
		\mathbb{E} \left(  | | \boldsymbol{X} - \boldsymbol{P} \boldsymbol{T} ^ { \top } -\boldsymbol{\bar{X}} | | _ { F } ^ { 2 } \right) &= | | \boldsymbol{X} - \mathbb{E}(\boldsymbol{P}) \mathbb{E} (\boldsymbol{T}) ^ { \top } - \mathbb{E}( \boldsymbol{\bar{X}_{ji}}) | | _ { F } ^ { 2 } 
		+\operatorname { t r } \left(  n \mathbb{E} (\boldsymbol{P})^{\top}\mathbb{E} (\boldsymbol{P}) \boldsymbol{\Sigma} ^ { \boldsymbol{t}_i } \right)  \\
		&+ \operatorname { t r } \left(  m \mathbb{E} (\boldsymbol{T})^{\top}\mathbb{E} (\boldsymbol{T}) \boldsymbol{\Sigma} ^ { \boldsymbol{p}_i } \right) 
		+\operatorname { t r } \left(  mn \boldsymbol{\Sigma} ^ { \boldsymbol{p}_i }\boldsymbol{\Sigma} ^ { \boldsymbol{t}_i } \right) +\sum_{i=1}^{n}\sum_{j=1}^{m}\boldsymbol{\Sigma}^{\boldsymbol{\bar{X}}}_{j i} .
	\end{split}
\end{align}

\subsection{Variational inference for VBSPCA with Laplace prior}

Following \eqref{24}, the latent variables and parameters of the Laplace distribution-based VBSPCA are $\boldsymbol{\Theta} = \left(  \boldsymbol{P},\boldsymbol{t}, \boldsymbol{\bar{x}}, \boldsymbol{\vartheta}, \boldsymbol{\eta} \right) $. Furthermore, $\boldsymbol{\Theta}$ can be written as 
\begin{align}
	Q(\boldsymbol{\Theta}) =q( \boldsymbol{P}) q(\boldsymbol{t})  q(\boldsymbol{\bar{x}})    q(\boldsymbol{\vartheta})    q(\boldsymbol{\eta}).
\end{align}


Optimizing the approximate distribution $Q(\boldsymbol{\Theta})$  based on the joint distribution of \eqref{24} gives the form of these parameters and latent variables
\begin{subequations}
	\begin{align}
		\label{e} q( \boldsymbol{P}) &=\prod _ { i = 1 } ^ { r } \mathcal{N} ( \boldsymbol{P} _ { i  } , | \boldsymbol{\mu}^{\boldsymbol{P}} _ { i  } , \boldsymbol{\Sigma}^{\boldsymbol{P}}_ { i  } )  \\
	\label{f}	q( \boldsymbol{t}) &=\prod _ { i = 1 } ^ { n } \mathcal{N} ( \boldsymbol{t} _ { i  } , | \boldsymbol{\mu}^{\boldsymbol{t}} _ { i  } , \boldsymbol{\Sigma}^{\boldsymbol{t}}_ { i  } )  \\
	\label{g}	q( \boldsymbol{\bar{x}}) &=\mathcal{N} ( \boldsymbol{\bar{x}} , | \boldsymbol{\mu}^{\boldsymbol{\bar{x}}}  , \boldsymbol{\Sigma}^{\boldsymbol{\bar{x}}} )  \\
	\label{h}	q( \boldsymbol{\vartheta}) &= \text{Gam}(\boldsymbol{\vartheta}| c,d)  \\
	\label{i}	q(\boldsymbol{\eta}) &= \frac { 1 } { \sqrt { \pi \boldsymbol{\eta} _ { i  j } \varphi} } \exp \left(  - \frac { 1 } { 2 \boldsymbol{\eta} _ { ji } } ( \boldsymbol{P} _ { j i } ) ^ { \top } ( \boldsymbol{P} _ { j i } ) - \frac{\boldsymbol{\eta}_{j i}}{\varphi}+ \sqrt{\frac{2}{\varphi}} | \boldsymbol{P}_{j i} |    \right) 
	\end{align}
\end{subequations}

The approximate distribution $q( \boldsymbol{P})$ in\eqref{e} is given by
\begin{align}
	\begin{split}
		{ \boldsymbol{\mu}^{\boldsymbol{P}} _ { i  }}  & =   \mathbb{E}(\boldsymbol{\vartheta}) \boldsymbol{\Sigma}^{\boldsymbol{P}}_ { i  } \sum_{i=1}^{n}\mathbb{E}(\boldsymbol{x}_i)
		(\boldsymbol{y}_i-\mathbb{E}(\boldsymbol{\bar{x}}))
		\\
		\boldsymbol{\Sigma}^{\boldsymbol{P}}_ { i  } & = \left(  \mathbb{E}(\boldsymbol{\vartheta})  \sum_{i=1}^{n}\mathbb{E}(\boldsymbol{x}_i\boldsymbol{x}_i^{\top}) \text{dg}(\mathbb{E}(\boldsymbol{\eta}_{i})) + \boldsymbol{I} \right)  ^ { - 1 } \text{dg}(\mathbb{E}(\boldsymbol{\eta}_{i})) .
	\end{split}
\end{align}
where $\text{dg}$ converts the vector to a diagonal matrix.

The approximate distribution $q( \boldsymbol{t})$ in \eqref{f} is given by
\begin{align}
	\begin{split}
		{ \boldsymbol{\mu}^{\boldsymbol{t}} _ { i  }}  & =   \mathbb{E}(\boldsymbol{\vartheta}) 	\boldsymbol{\Sigma}^{\boldsymbol{t}}_ { i  } \mathbb{E}({\mathbb{\boldsymbol{P}}})(\boldsymbol{y}_i-\mathbb{E}(\boldsymbol{\bar{x}}))
		\\
		\boldsymbol{\Sigma}^{\boldsymbol{t}}_ { i  } & = \left(  \mathbb{E}(\boldsymbol{\vartheta})  \mathbb{E} (\boldsymbol{P} ^ { \top } \boldsymbol{P})  + \boldsymbol{I} \right)  ^ { - 1 }  .
	\end{split}
\end{align}

The approximate distribution $q( \boldsymbol{\bar{x}})$ in \eqref{g} is given by
\begin{align}
	\begin{split}
		\boldsymbol{\mu}^{\boldsymbol{\bar{x}}}   & =   \mathbb{E}(\boldsymbol{\vartheta}) 	\boldsymbol{\Sigma}^{\boldsymbol{\bar{x}}} 
		\sum_{i=1}^{n}\left(  \boldsymbol{y}_i - \mathbb{E}(\boldsymbol{P}) \mathbb{E}(\boldsymbol{\boldsymbol{x}_i})     \right) 
		\\
		\boldsymbol{\Sigma}^{\boldsymbol{\bar{x}}} & = \left(  n\mathbb{E}(\boldsymbol{\vartheta})  + \varsigma \right)  ^ { - 1 }  \boldsymbol{I} .
	\end{split}
\end{align}

The approximate distribution $q( \boldsymbol{\vartheta})$ in \eqref{h} is given by
\begin{align}
	\begin{split}
		c   & =   c_0+\frac{nm}{2}
		\\
		d & = d_0+ \frac{1}{2} \sum_{i=1}^{n} {|| \boldsymbol{\bar{x}} ||}^{2} +tr( \mathbb{E}(\boldsymbol{P}^{\top} \boldsymbol{P}) \mathbb{E}(\boldsymbol{x}_i \boldsymbol{x}_i)^{\top} ) \\
		&+2\mathbb{E}(\boldsymbol{\bar{x}})^{\top}\mathbb{E}(\boldsymbol{P})\mathbb{E}(\boldsymbol{x}_i) -2\boldsymbol{y}_i^{\top}\mathbb{E}(\boldsymbol{P})\mathbb{E}(\boldsymbol{x}_i) - 2{\boldsymbol{y}_i}^{\top}\mathbb{E}(\boldsymbol{\bar{x}})
		+|| \boldsymbol{y}_i ||^{2}.
	\end{split}
\end{align}

The approximate distribution $q( \boldsymbol{\eta})$ \eqref{i} is given by
\begin{align}
	\mu_{\boldsymbol{\eta}_{ji}} = \frac{1}{2} (\varphi+\sqrt{2\varphi} | \boldsymbol{P}_{ji}| ).
\end{align}

\section{Dynamic Process Monitoring}  \label{sec:fault detection}
\subsection{Fault detection}

In this section, fault detection is accomplished and the dynamics of the latent variables are examined after the projection loading matrix $\boldsymbol{P}$ is calculated by variational inference using the VBSPCA approach, which is based on Gaussian prior and Laplace prior. 

\begin{remark}
	Fault detection and diagnosis depends more on the dynamics of the latent variables $\boldsymbol{t}^{\ast}_k $ of the newly collected samples, although it also depends on the projection matrix $ \boldsymbol{P}$.
	Therefore, it is more interpretable to analyze the dynamics of the latent variables directly.
\end{remark}

A new collected sample $\boldsymbol{y}_{k} \in \mathbb{R}^{m } $ is obtained at time $k$ with time interval $L$, the latent variable  $\boldsymbol{t}^{\ast}_k$ of the new collected sample $\boldsymbol{y}_{k}$ is obtained by a linear transformation of $\boldsymbol{P}$
\begin{align}
	\boldsymbol{t}^{\ast}_k = \boldsymbol{P}^{\top}\boldsymbol{y}_{k} 
\end{align}
where $\boldsymbol{t}^{\ast}_k$ is a potential time series, which means that $\boldsymbol{t}^{\ast}_k$ at the current moment is influenced by the previous moment.

As the dynamic relationship of the latent variable $\boldsymbol{t}^{\ast}_k$ is not explained, it is necessary to build a dynamic model to characterize the auto correlation.
As a result, $\boldsymbol{t}^{\ast}_k$ can be expressed as the following VAR model
\begin{align}
	\boldsymbol{t}^{\ast}_k=\underset{d=1}{\overset{\tau}{\mathop \sum }}\,{\boldsymbol{\omega}_{d}}{{\boldsymbol{t}}^{\ast}_{k-d}}+{{\boldsymbol{\epsilon }}_{k}}, k = \tau+1,\cdots,L  
\end{align}
where $\tau$ is the order of dynamics,  $\boldsymbol{\epsilon}_k $ is zero mean Gaussian noise, ${\boldsymbol{\omega}_{i}} \in \mathbb{R}^{r \times r}, d = 1,2, \cdots, \tau$ represents the coefficient matrix of the VAR model.

The objective is to learn the coefficient matrix $\boldsymbol{\omega}$ from the latent variable, and the new linear expression for $\boldsymbol{\omega}$ and ${{\boldsymbol{\chi}}_{k}}$ is rewritten as
\begin{align}
	\begin{split}
		\boldsymbol{\omega} &= \left[{\boldsymbol{\omega}_{1}},\cdots,{\boldsymbol{\omega}_{\tau}} \right] ^{\top} \in \mathbb{R}^{(r \cdot \tau)\times r} \\
		{{\boldsymbol{\chi}}_{k}} &= \left[{ \boldsymbol{t}^{\ast}_{k-1} }^{\top},\cdots, {\boldsymbol{t}^{\ast}_{k-\tau}}^{\top} \right]  \in \mathbb{R}^{(r \cdot \tau)}
	\end{split}
\end{align}

The VAR model for the latent variable $\boldsymbol{t}$ can be rewritten as
\begin{align}
	{{\boldsymbol{t}}_{k}^{\ast}}\approx \underset{d=1}{\overset{\tau}{\mathop \sum }}\,{{\boldsymbol{\omega}}_{k}}{{\boldsymbol{t}}_{k-d}^{\ast}}={\boldsymbol{\omega}}^{\top}{{\boldsymbol{\chi}}_{k}} , k = \tau+1,\cdots,L  
\end{align}

In high dimensional scaling, where the transition matrix driving the temporal dynamics displays a more complex structure. In order to train the sparse projection in the regression model, $\ell_1$ regularization is therefore required
\begin{align}
	\boldsymbol{\hat{\omega}}=\underbrace{\arg \min}_{\boldsymbol{\omega}}\frac{1}{2} {\|\boldsymbol{Z}-\boldsymbol{Q}\boldsymbol{\omega}\|_{2}^{2}}+\lambda {{\left\| \boldsymbol{\omega} \right\|}_{1}}
\end{align}
%
where $ \boldsymbol{Z} $ and $ \boldsymbol{Q} $ are defined as $\boldsymbol{Z}=\left[ \boldsymbol{t}_{\tau+1}^{\top},...,\boldsymbol{t}_{L}^{\top} \right]^{\top} \in \mathbb{R}^{(F-\tau)\times r}$, $ \boldsymbol{Q} ={{\left[ {{\boldsymbol{\chi}}_{\tau+1}^{\top}},...,{{\boldsymbol{\chi}}_{L}^{\top}} \right]^{\top} \in \mathbb{R}^{(F-\tau)\times(r\cdot \tau)} }}$, respectively.

\begin{remark}
	A sparse VAR about $\boldsymbol{\omega}$ is built such that $\boldsymbol{\omega}$ becomes a sparse matrix for feature selection on the latent variable $\boldsymbol{t}^{\ast}_k$ of the newly collected sample $\boldsymbol{y}_k$. This further reduces the noise perturbation of the newly collected sample $\boldsymbol{y}_k$ and improves the model interpretability.
\end{remark}

The prediction of $\hat{\boldsymbol{t}}_k$ from the sparse VAR model is 
\begin{align}
	\hat{\boldsymbol{t}}_k = \sum _{d=1}^{\tau}\boldsymbol{\hat{\omega}}_d	\boldsymbol{t}^{\ast}_{k-d} = \sum _{d=1}^{\tau}\boldsymbol{\hat{\omega}} _i \boldsymbol {P}^{\top}{\boldsymbol y}_{k-d} , k = \tau+1,\cdots,L   .
\end{align}

From the predicted latent variable $\hat{\boldsymbol{t}}_k$, the $T_2$ statistics and $SPE$ statistics can be obtained, which are the essential indicators for performing fault detection.

The $T_2$ statistic can be constructed as
\begin{align}
	T_2 = \hat{\boldsymbol{t}}_k \boldsymbol{\Lambda}\hat{\boldsymbol{t}}_k ^{\top}.
\end{align}
where $\boldsymbol{\Lambda}$ is a diagonal matrix.

The $SPE$ statistic can be constructed as
\begin{align}
	SPE = {{\left\| \left( \boldsymbol{I}-\boldsymbol{P}{{\boldsymbol{P}}^{\top }} \right)\hat{\boldsymbol{t}}_k \right\|}^{2}}.
\end{align}

In this work, the kernel density estimation (KDE) method is used to calculate the control limit, which is a non-parametric method to build an unknown density function

\begin{align} \label{17}
	{{{\mathop{\hat{f}}}\,}_{h}}\left( e \right)=\frac{1}{L}\underset{i=1}{\overset{L}{\mathop{\sum }}}\,{{K}_{h}}\left( e-{{e}_{i}} \right)=\frac{1}{Lh}\underset{i=1}{\overset{L}{\mathop{\sum }}}\,{{K}_{h}}\left( \frac{e-{{e}_{i}}}{h} \right).
\end{align}
where $ {{K}_{h}} $ is a non-negative kernel function and $ h $ is a relative smoothing parameter called bandwidth.  $ {e} $ is the reconstruction error, to construct the density estimate accordingly. The confidence limit can be obtained with a confidence level $ \alpha = 0.95 $.
When the statistic exceeds the confidence limit, the test sample $ \boldsymbol{y}_{k} $ is considered faulty.

\subsection{Fault diagnosis}

A detectable fault should have an impact on the measurement vector that deviates from the normal scenario. 
The measurement vector for the fault-free part is denoted by $\boldsymbol{\psi}$ and will be unavailable when a fault occurs\cite{alcalaReconstructionbasedContributionProcess2009}. 
In the presence of a process fault, ${\boldsymbol{y}}$ is represented as
\begin{align}
	\boldsymbol{y}=\boldsymbol{\psi}+\boldsymbol{f} \boldsymbol{\phi }
\end{align}
where $\boldsymbol{\phi }$ is orthonormal so that $ ||\boldsymbol{f}||$ represents the magnitude of the fault.
$ \boldsymbol {\psi} $ is the fault-free part of the measurement.

The objective of fault reconstruction is to estimate the fault-free part $\boldsymbol{\psi}$ by removing the effects of faults. 
A reconstructed sample vector $\boldsymbol{\psi}_k \in \boldsymbol{\psi}$ is calculated by correcting the effect of a fault on the process data $\boldsymbol{y}$
\begin{align}
	{\boldsymbol{\psi}_{k}}=\boldsymbol{y}-{f}_k {{\phi }_{k}} 
\end{align}

The fault detection index for reconstructing variables can be written as
\begin{align} \label{54} 
	\text{Index}\left( {{\boldsymbol{\psi}}_{k}} \right)=\boldsymbol{y}_{k}^{\top}\boldsymbol{\Phi} {{\boldsymbol{y}}_{k}}=\|{{\boldsymbol{y}}_{k}}\|_{\boldsymbol{\Phi} }^{2}=\|\boldsymbol{\psi}-{{f}_{k}}{{\phi }_{k}}\|_{\boldsymbol{\Phi}}^{2} .
\end{align}

The purpose of reconstruction is to find the value that minimizes $ {f}_{k} $. 
	Taking the derivation of \eqref{54}, 
${{f}_{k}}={{\left( { \phi}_{k}^{\top}{\boldsymbol{\Phi}}{{ \phi}_{k}} \right)}^{-1}}{ \phi}_{k}^{\top}{\boldsymbol{\Phi}}{\boldsymbol{y}}$ can be obtained. 
The reconstruction-based contribution $ \text{RBC}_{k}^{\text{Index}} $ along the fault direction can be expressed as
\begin{align}
	\text{RBC}_{k}^{\text{Index}}=\|{{ \phi}_{k}}{{f}_{k}}\|_{\boldsymbol{y}}^{2}
	={\boldsymbol{\psi}}^{\top}{\boldsymbol{\Phi}}{{ \phi}_{k}}{{\left( { \phi}_{k}^{\top}{\boldsymbol{\Phi}}{{ \phi}_{k}} \right)}^{-1}}{ \phi}_{k}^{\top}{\boldsymbol{\Phi}}\boldsymbol{y} .
\end{align}

The reconstructed index $\text{Index}(\boldsymbol{\psi}_k )$ in \eqref{54} can be expressed as
\begin{align}
	\text{Index}(\boldsymbol{\psi}_k )= \boldsymbol{y}^{\top} \boldsymbol{\Phi}  \left[ \boldsymbol{I} -  { \phi}_{k} {\left( { \phi}_{k}^{\top}{\boldsymbol{\Phi}}{{ \phi}_{k}} \right)} ^{-1} { \phi}_{k}^{\top} \boldsymbol{\Phi} \right]    \boldsymbol{y}
	=\boldsymbol{y}^{\top} \boldsymbol{\Phi}\boldsymbol{y}-\text{RBC}_{k}^{\text{Index}}
\end{align}

As a result, $\text{Index}(\boldsymbol{x})=\text{Index}(\boldsymbol{\psi}_k )+\text{RBC}_{k}^{\text{Index}}$.

For $T_2$ indexes, $\text{RBC}_{k}^{T_2}$ can be expressed as
\begin{align}
	\text{RBC}_{k}^{T_2}=\frac{ (  { \phi}_{k}^{\top}  \boldsymbol{P}\boldsymbol{\Lambda}^{-1}\boldsymbol{P}^{\top}   \boldsymbol{y})^2 }{  {\left( \boldsymbol{P}\boldsymbol{\Lambda}^{-1}\boldsymbol{P}^{\top}\right) }_{kk} }
\end{align}
where ${  {\left( \boldsymbol{P}\boldsymbol{\Lambda}^{-1}\boldsymbol{P}^{\top}\right) }_{kk} }$ is the $i$th diagonal element of ${  {\left( \boldsymbol{P}\boldsymbol{\Lambda}^{-1}\boldsymbol{P}^{\top}\right) } }$.

For $SPE$ indexes, $\text{RBC}_{k}^{SPE}$ can be expressed as
\begin{align}
	\text{RBC}_{k}^{SPE}=\frac{ (  { \phi}_{k}^{\top}  \left( \boldsymbol{I}-\boldsymbol{P}^{\top}\boldsymbol{P} \right)  \boldsymbol{y})^2 }{{\left( \boldsymbol{I}-\boldsymbol{P}^{\top}\boldsymbol{P} \right)}_{kk} }
\end{align}
where ${{\left( \boldsymbol{I}-\boldsymbol{P}^{\top}\boldsymbol{P} \right)}_{kk} }$ is the $i$th diagonal element of ${{\left( \boldsymbol{I}-\boldsymbol{P}^{\top}\boldsymbol{P} \right)} }$.

\section{Industrial AWE hydrogen production studies} \label{sec:experiments}

\begin{figure}[H]
		\centering
		\includegraphics[width=7in]{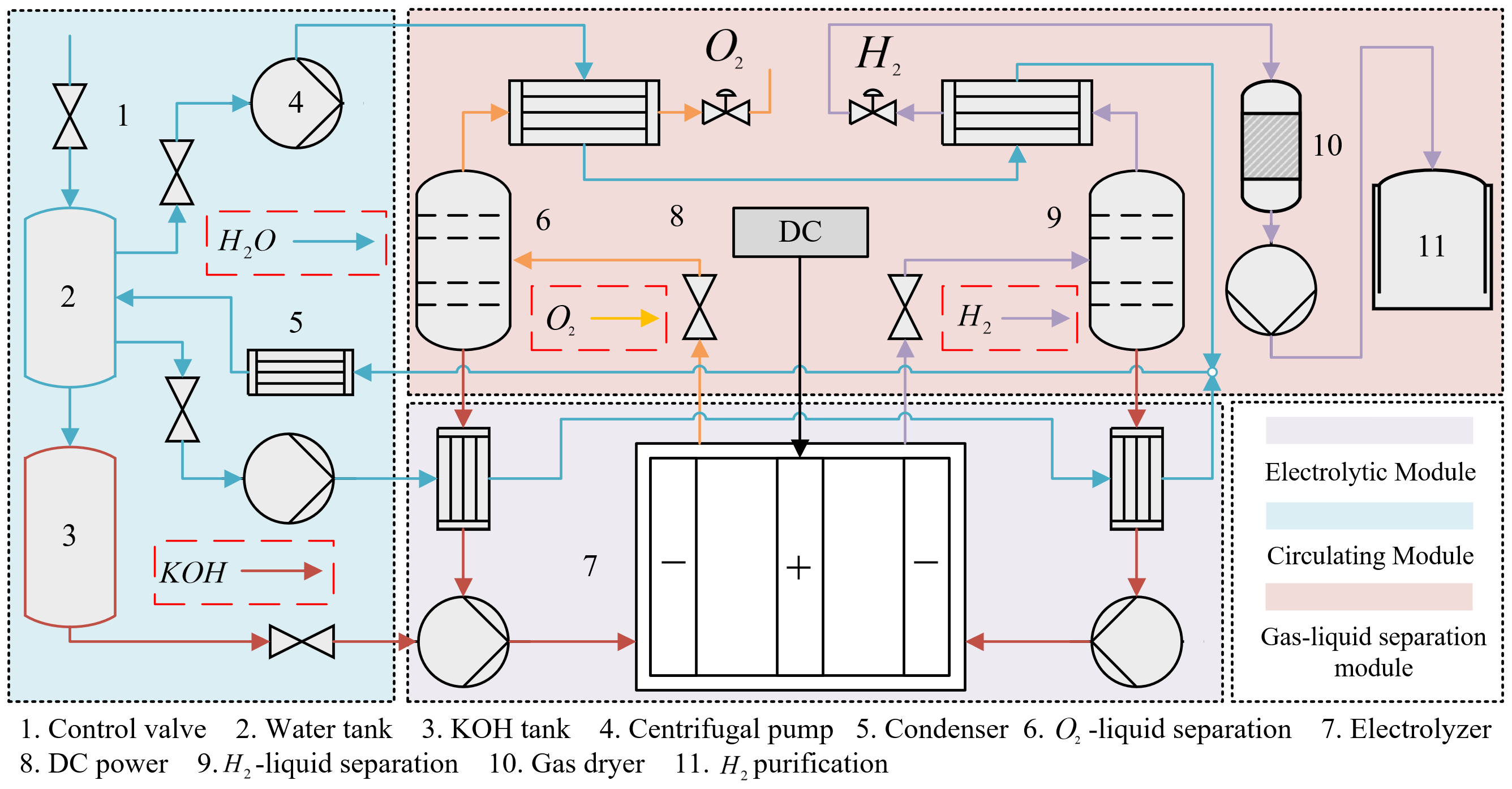}
		\caption{ Flow chart of alkaline electrolysis system.}
		\label{fig.1}
		\end{figure}

	The AWE hydrogen production industrial process consists of three main parts, electrolysis module, gas-liquid separation module, and circulation module. As shown in Figure \ref{fig.1}, they include alkaline electrolyzer, heat exchanger, gas-liquid separator, condenser and dryer.
		The direct current supply powers the alkaline water electrolyzer, and the electrolyte is then decomposed to generate hydrogen and oxygen.
	Hydrogen and oxygen are dissolved in the electrolyte. When the pressure rises to the rated value, the mixtures is sent to the gas-liquid separator through the back pressure valve, where gas and electrolyte are separated.
	After separation, the gases and liquid go through different routes.
	The filtered electrolyte is cooled to the working temperature by the condenser, and then the electrolyte is returned to the electrolytic cell through the circulating pump, which can reduce the temperature of the electrolyzer.
	The separated gas is cooled by the condenser, and the cooled gas enters the gas dryer for dehydration. In order to improve the purity of hydrogen, hydrogen is purified and stored by $ H_2 $ purification equipment.
	The liquid flows through the circulation pump and filter and then returns to the electrolysis cell from the bottom of the separators. The hydrogen is tested for purity and then collected and stored.
	During the electrolysis process, the circulation module cools the system while circulating the electrolyte. Pure water is consumed by electrolysis, so it is necessary to supplement water to get electrolyte. The generated electrolyte and the separated electrolyte are circulated into the electrolyzer to maintain the set concentration.
	
	The data of hydrogen production process used in this study comes from the real data of a large-scale hydrogen production plant, including normal data and fault data. 
	According to privacy requirements of the plant, the raw data is desensitized in order to ensure the privacy and security of the information.
 The variables of the electrolytic water hydrogen production process can be divided into control variables and process variables, and the regulation and monitoring of the variables can be realized by the automated control system to ensure the stability, efficiency and safety of the electrolytic water hydrogen production process. The plant collects 32 variables, including 10 control variables and 22 process variables. Table \ref{Table1} lists the control variables and Table \ref{Table2} lists the process variables.
	The main faults that may occur in a AWE hydrogen production system are summarized in the Table \ref{Table3}.
Ten common faults are present in all three components of the AWE hydrogen production system.

\begin{table}[H]\centering
	\caption{Control variable in AWE system.}
	\label{Table1}
	\setlength{\tabcolsep}{1mm}
	\begin{tabular}{c c c} 
		\toprule
		Control variable & Variable description & Unit \\
		\hline
		CV(1) & Hydrogen mass flow  &	$ km^3/h $	\\
		CV(2) &	 Oxygen mass flow	&	$ km^3/h $  \\
		CV(3) & Hydrogen gas-liquid separator liquid level	&	L  \\
		CV(4) & Oxygen gas-liquid separator liquid level  &  L   \\ 
		CV(5) & Water tank return flow & $ km^3/h $  \\ 		
		CV(6) & KOH tank outlet flow & $ km^3/h $  \\ 	
		CV(7) & Electrolytic cell electrode current & A  \\ 	
		CV(8) & Electrolyzer total voltage & V \\ 	
		CV(9) & Electrolyte concentration & wt\%  \\ 	
		CV(10) & Electrolyzer inlet flow & $ km^3/h $  \\
		\bottomrule
	\end{tabular}
\end{table}

\begin{table}[H]\centering
	\caption{Process variable in AWE system.}
	\label{Table2}
	\setlength{\tabcolsep}{1mm}
	\begin{tabular}{c c c} 
		\toprule
		Process variable & Variable description & Unit \\
		\hline
		PV(1) & Electrolyzer cell temperature  &	$^{\circ}$C 	\\
		PV(2) &	 Electrolyzer temperature	&	$^{\circ}$C  \\
		PV(3) & Electrolyte level	&	L  \\
		PV(4) & Electrolyzer pressure  & kPa   \\ 
		PV(5) & Hydrogen purity & umol/mol  \\ 		
		PV(6) & Oxygen purity & umol/mol  \\ 	
		PV(7) & Hydrogen separator pressure & kPa  \\ 	
		PV(8) & Oxygen separator pressure & $^{\circ}$C \\ 	
		PV(9) & Hydrogen condenser temperature & $^{\circ}$C \\ 	
		PV(10) & Oxygen condenser temperature & $^{\circ}$C  \\
		PV(11) & Oxygen electrolyte condenser temperature & $^{\circ}$C  \\
		PV(12) & Hydrogen electrolyte condenser temperature & $^{\circ}$C  \\
		PV(13) & Return water condenser temperature & $^{\circ}$C  \\
		PV(14) & KOH tank level & L  \\
		PV(15) & Water tank level & L  \\
		PV(16) & Gas dryer humidity &  \%RH \\
		PV(17) & Gas dryer pressure & kPa \\
		PV(18) & Return pipeline pressure & kPa \\
		PV(19) & Hydrogen separator electrolyte concentration & wt\% \\
		PV(20) & Oxygen separator electrolyte concentration & wt\% \\
		PV(21) & Hydrogen purifier pressure & kPa\\
		PV(22) & Hydrogen purifier purity & umol/mol\\
		\bottomrule
	\end{tabular}
\end{table}

	\begin{table}[H]\centering
	\caption{Summary of different type of faults in AWE system.}
	\label{Table3}
	\setlength{\tabcolsep}{5mm}
	\begin{tabular}{c c c} 
		\toprule
		Tag & Fault description & Module \\
		\hline
		1 & Electrolysis cell voltage rise &	Electrolytic module	\\
		2 &	 Electrolyte inter-electrode voltage is abnormal	&	Electrolytic module \\
		3 & Electrolyzer short circuit	&	Electrolytic module  \\
		4 & Reduced gas purity  & Gas-liquid separation module \\ 
		5 & Increased hydrogen oxygen liquid level difference & Gas-liquid separation module\\ 		
		6 & A sharp rise or fall in the hydrogen or oxygen level & Gas-liquid separation module\\ 	
		7 & Unstable pressure & Gas-liquid separation  \\ 	
		8 & Electrolyte stops circulating & Circulating module\\ 	
		9 & Increased electrolyzer temperature & Circulating module \\ 	
		10 & Water tank bubbles & Circulating module  \\
		\bottomrule
	\end{tabular}
\end{table}

	In this section, two faults in the three modules of the hydrogen production process are analyzed. All faults in the test sample were introduced from the 201st sample.
	Fault 2 is the  electrolyte inter-electrode voltage is abnormal. 
	The hydrogen production rate in the alkaline water electrolyzer depends on the DC power supply output. The higher the voltage applied to the water, the faster the water molecules will split into hydrogen and oxygen gases. This is because the voltage provides enough energy to overcome the bond energy of water and drive the electrolysis reaction. 
	When the voltage increases, more electrons are transferred to the water, making the reaction more vigorous. However, too high a voltage can also generate excessive bubbles that can lower the reaction efficiency. 
	Therefore, it is important to choose an optimal voltage for water electrolyzer to achieve the best balance between reaction speed and effect.
	When the voltage is abnormal, it can cause the electrolysis reaction to not proceed properly. High voltage can cause bubbles to accumulate in the electrolyzer, posing a fire and explosion hazard. Voltage fluctuations can lead to unstable electrolysis reactions and the hydrogen production system poses an explosion risk.

\begin{figure}[H]
		\centering
		\subfigure[Gaussian prior-based VBSPCA.]{
			\includegraphics[width=3.3in]{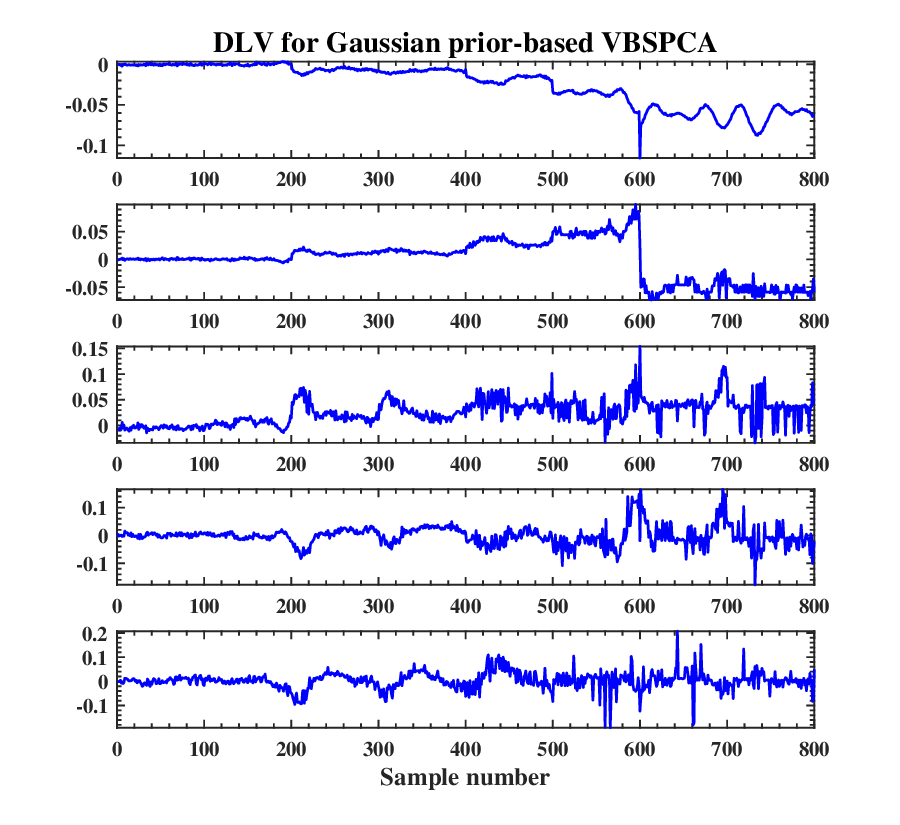}
		}
		\subfigure[Laplace prior-based VBSPCA.]{
			\includegraphics[width=3.3in]{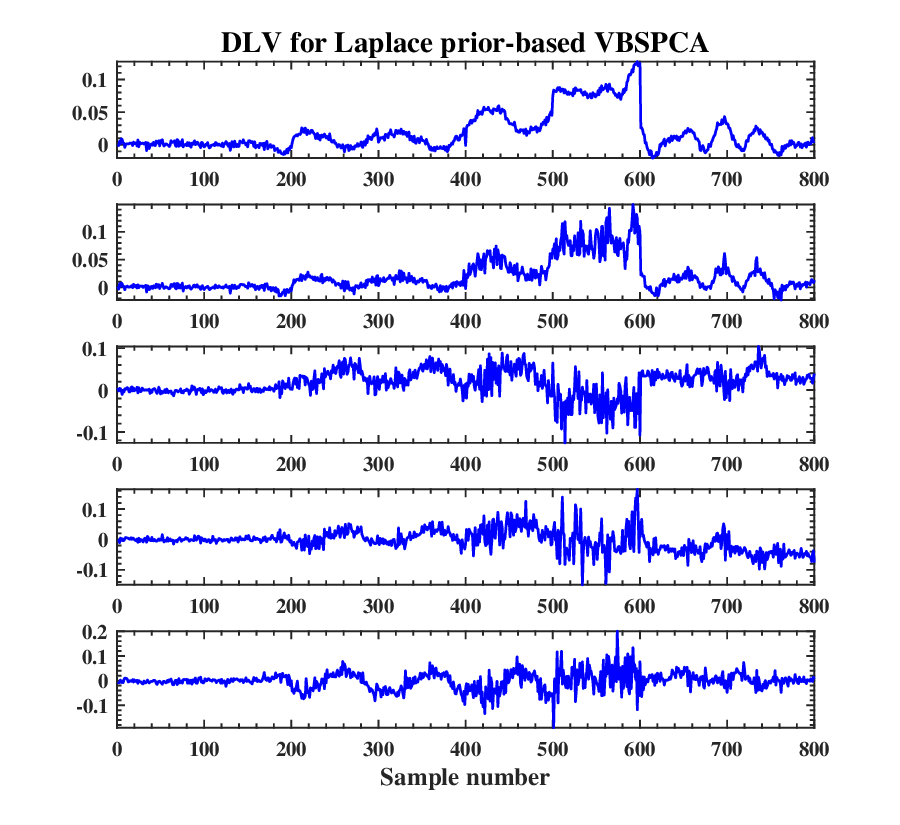}
		}
		\caption{Dynamic latent variables.}
		\label{fig.2}
	\end{figure}

\begin{figure}[H]
		\centering
		\subfigure[Gaussian prior-based VBSPCA.]{
			\includegraphics[width=3.3in]{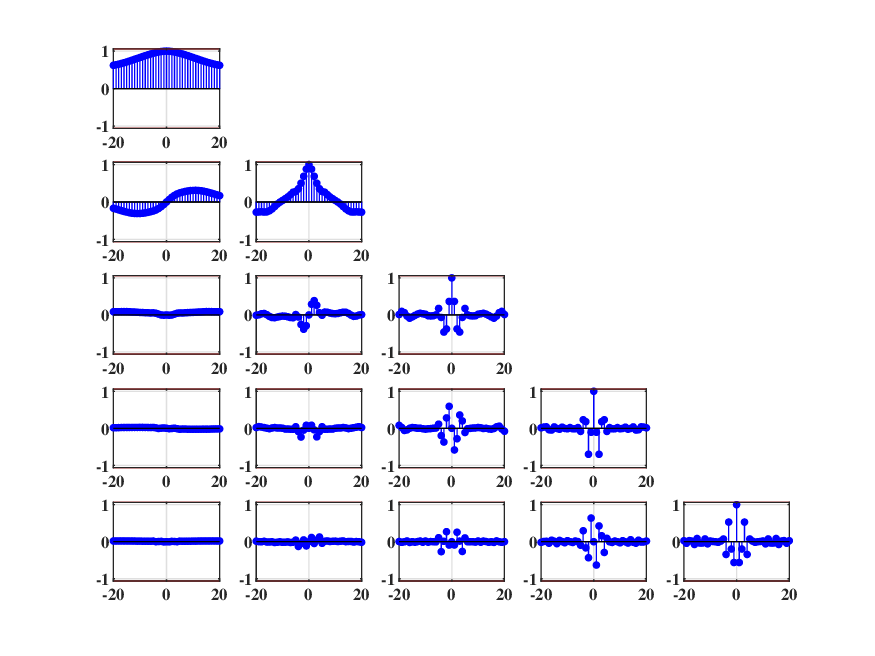}
		}
		\subfigure[Laplace prior-based VBSPCA.]{
			\includegraphics[width=3.3in]{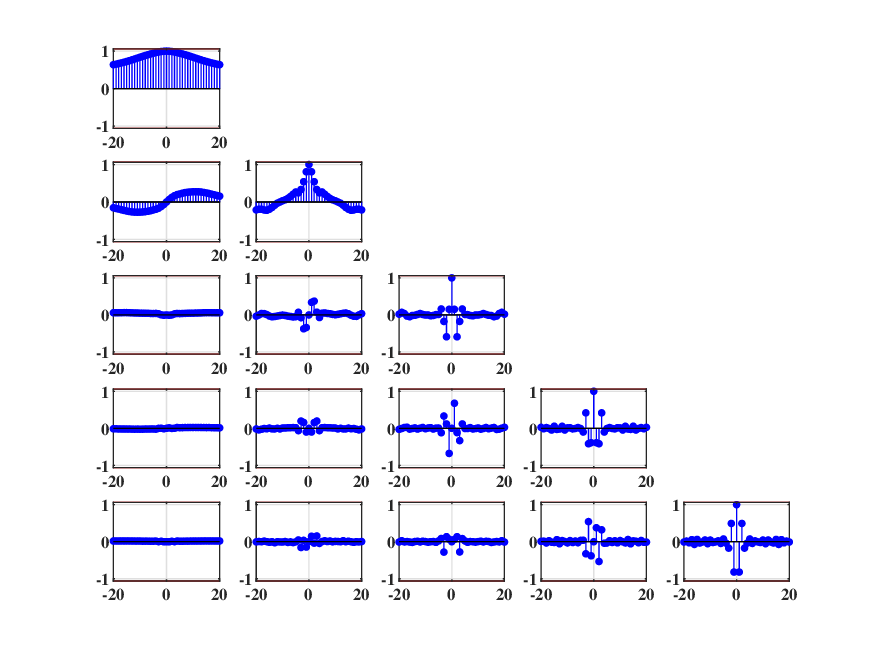}
		}
		\caption{Autocorrelation coefficients of dynamic latent variables.}
		\label{fig.3}
	\end{figure}

	By capturing the dynamic relationships of multiple variables through lagged values of each variable and regression analysis, the response of a dynamic latent variable to another variable at different points in time can be calculated. 
	The predicted values of the latent variables calculated by the sparse VAR model are shown in Figure \ref{fig.2}, where the five dynamic latent variables have been able to characterize the full range of changes in the system. 
	The correlations of the obtained dynamic latent variables were checked by plotting correlation plots. The correlations among the dynamic latent variables are presented in Figure \ref{fig.3}. As depicted, it is evident that the correlations have been eliminated through modeling, whereby the first five dynamic latent variables are extracted.

\begin{figure}[H]
	\centering
	\subfigure[Gaussian prior-based VBSPCA.]{
		\includegraphics[width=3.3in]{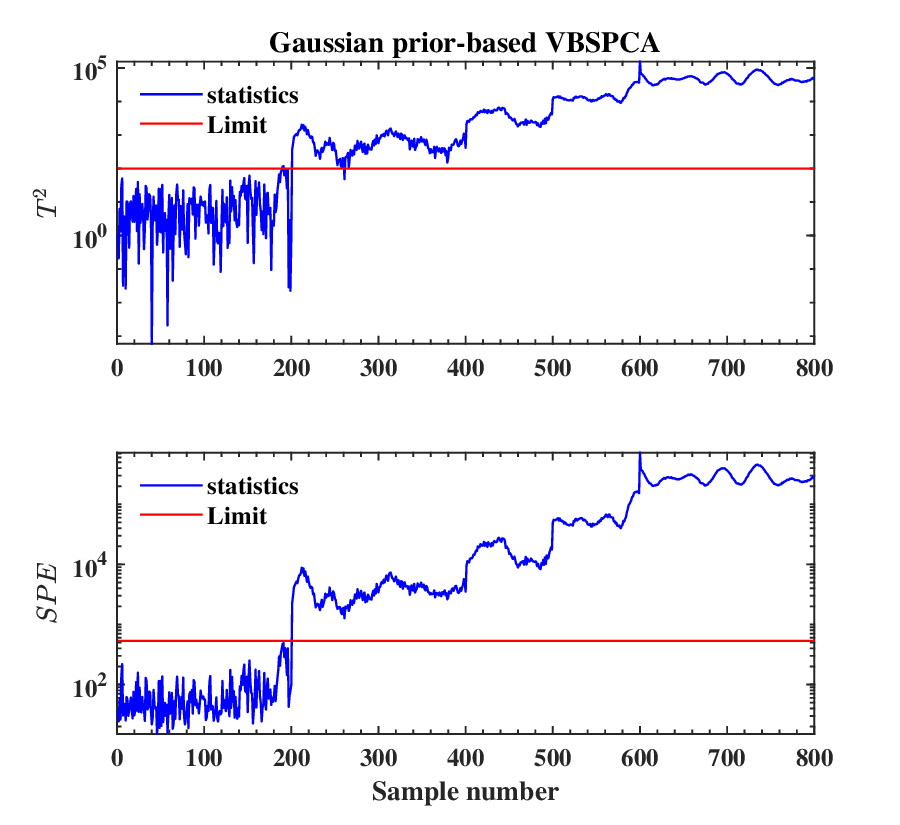}
	}
	\subfigure[Laplace prior-based VBSPCA.]{
		\includegraphics[width=3.3in]{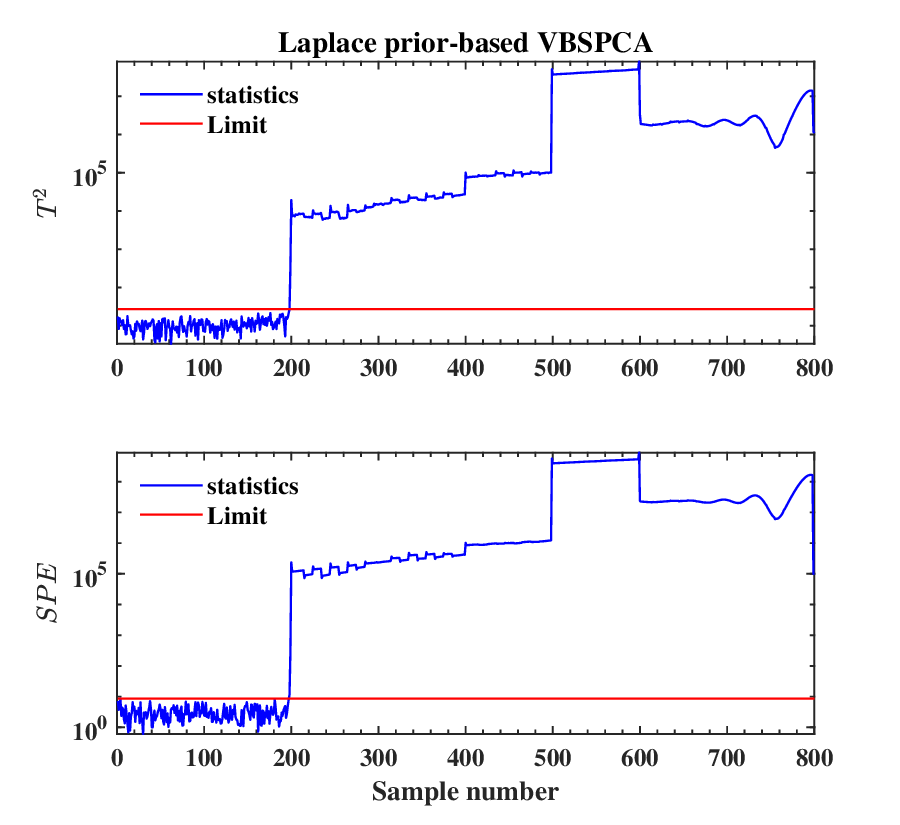}
	}
	\subfigure[Variational Bayesian dictionary learning.]{
		\includegraphics[width=3.3in]{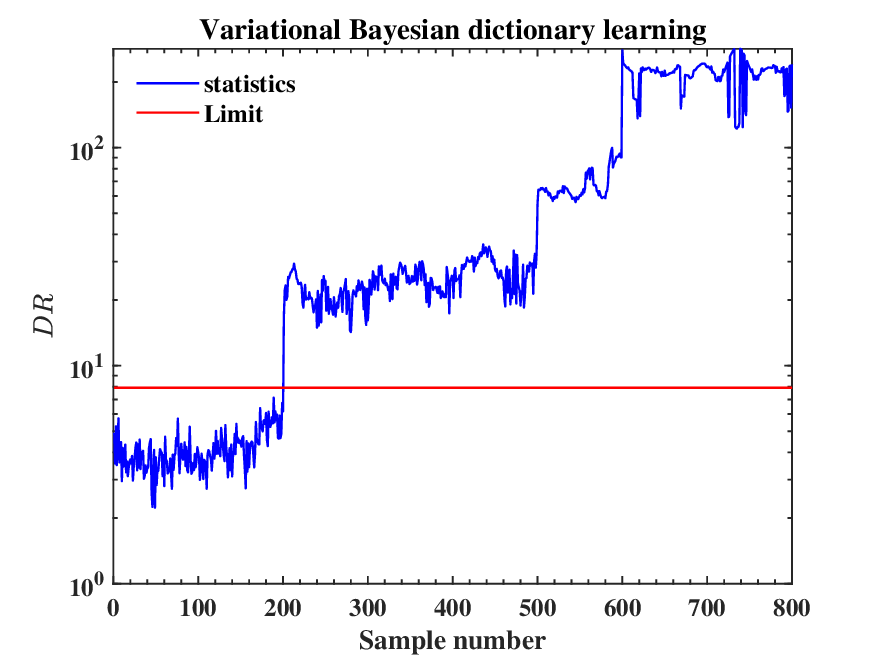}
	}
	\subfigure[Dynamic inner canonical correlation analysis.]{
		\includegraphics[width=3.3in]{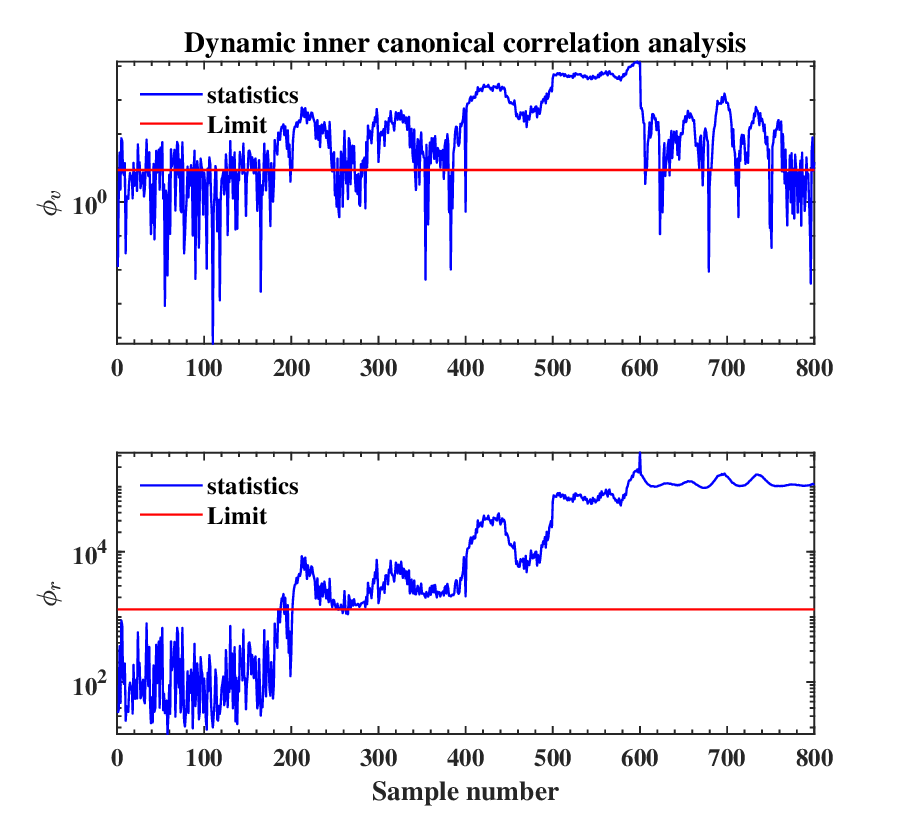}
	}
	\caption{Detection result of fault 2.}
	\label{fig.4}
\end{figure}

	The results of the fault detection are shown in Figure \ref{fig.4}, where the blue line represents the fault statistic, which presents the variation of the fault process. The red line is the confidence limit, and a statistic above the confidence limit indicates that a fault has been detected.
	In order to verify the advantages of the VBSPCA method, we compare the variational Bayesian dictionary learning (VBDL) method with the DiCCA method. 
	The VBDL obtains sparsity by constructing a dictionary matrix, but does not consider the dynamic relationships\cite{zhangVariationalBayesianDictionary2023a}. 
	The DiCCA method analyses the dynamic relationships of the latent variables but lacks robustness to noise perturbations. 
	 Fault 2 is heavily contaminated by noise, and three different Bayesian sparsity methods effectively detect the fault, but the DiCCA method does not detect all faults. 
	 In addition, the Laplace prior-based VBSPCA method has a better fault detection ability compared to the  Gaussian prior-based VBSPCA method.

\begin{figure}[H]
		\centering
		\subfigure[Gaussian prior-based VBSPCA with $T_2$.]{
			\includegraphics[width=3.3in]{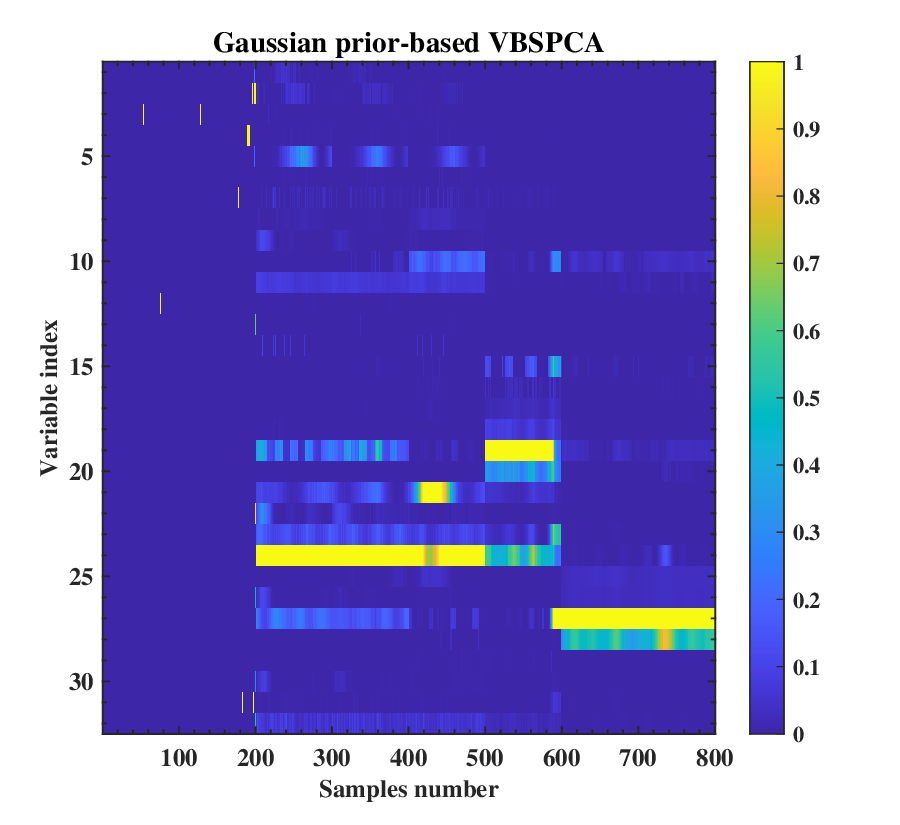}
		}
		\subfigure[Gaussian prior-based VBSPCA with $SPE$.]{
			\includegraphics[width=3.3in]{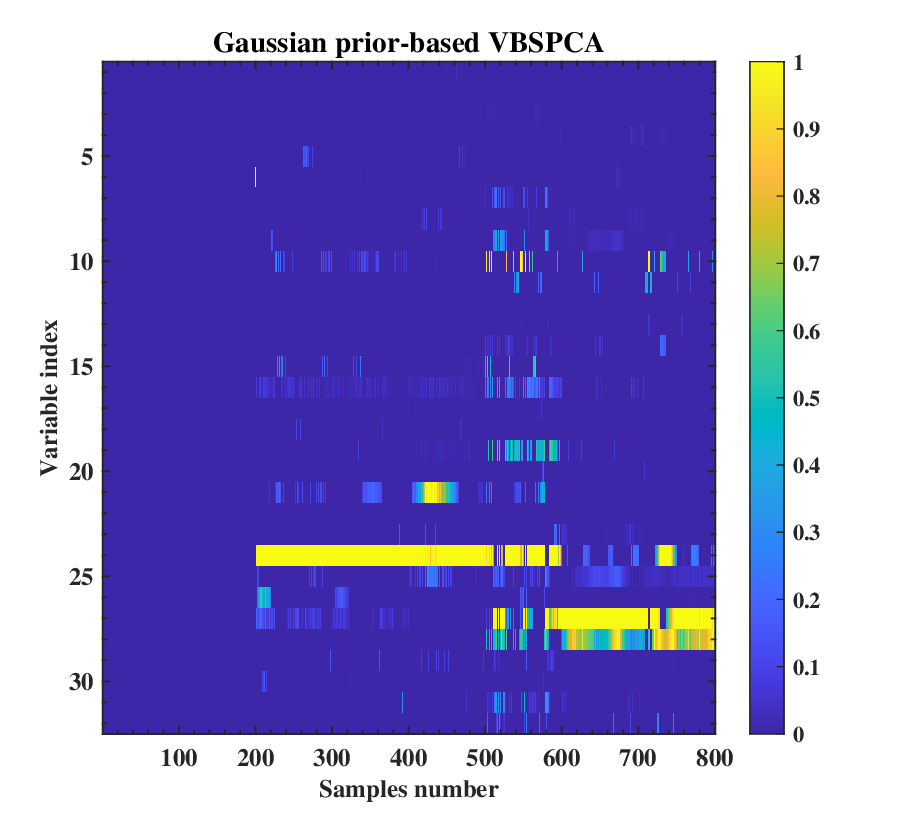}
		}
		\subfigure[Laplace prior-based VBSPCA with $T_2$.]{
			\includegraphics[width=3.3in]{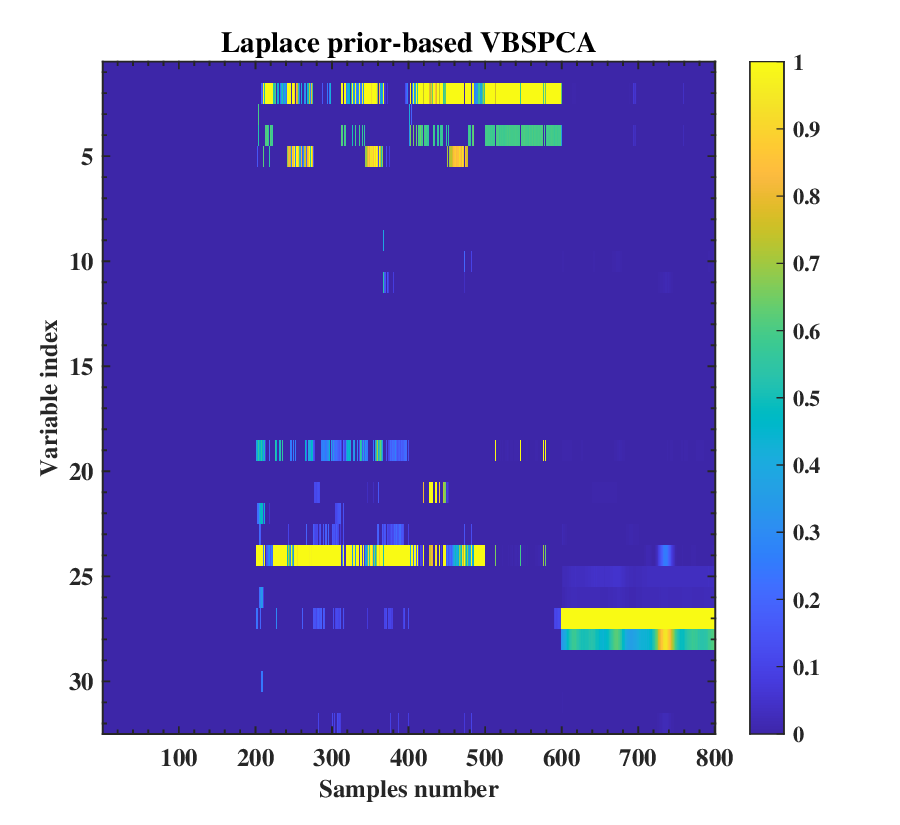}
		}
		\subfigure[Laplace prior-based VBSPCA with $SPE$.]{
			\includegraphics[width=3.3in]{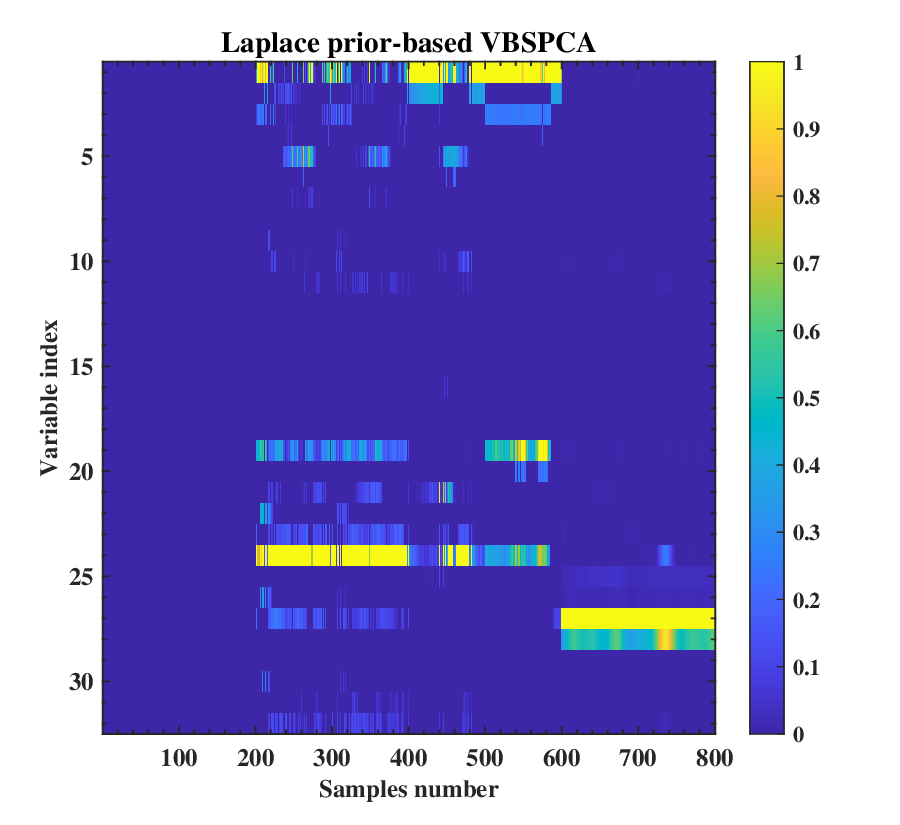}
		}
		\caption{Diagnosis result of fault 2.}
		\label{fig.5}
	\end{figure}

	The fault diagnosis results based on $T_2$ and $SPE$ indices are shown in Figures \ref{fig.5}, respectively. The horizontal axis represents the sample index, while the vertical axis represents the variable index. The RBC for each variable and sample is represented by the color depth, where deeper colors indicate higher RBC.
	The VBSPCA method based on both Gaussian prior and Laplace prior can diagnose the fault.
	The fault diagnosis results indicate that variable 24 plays a crucial role in the early stage of the fault, while variables 27 and 28 are the main factors in the later stage of the fault. 
	In addition, the Laplace prior-based VBSPCA suggests that the first five variables also contribute to the fault.
	
	Fault 6 is a sharp rise or fall in the hydrogen or oxygen level.
	If there is a sudden change in the hydrogen level, it may result in hydrogen leakage, which can cause a hydrogen explosion accident, leading to casualties and equipment damage. 
	There are several factors that can cause changes in the hydrogen level during the production process, including blockage of the balance valve or pipeline, and blockage of the outlet of the separation and defogging device. 
	Additionally, damage to the diaphragm inside the electrolytic cell due to long-term operation can also result in improper collection of hydrogen or oxygen, leading to a sudden change in the hydrogen or oxygen liquid level.

\begin{figure}[H]
	\centering
	\subfigure[Gaussian prior-based VBSPCA.]{
		\includegraphics[width=3.3in]{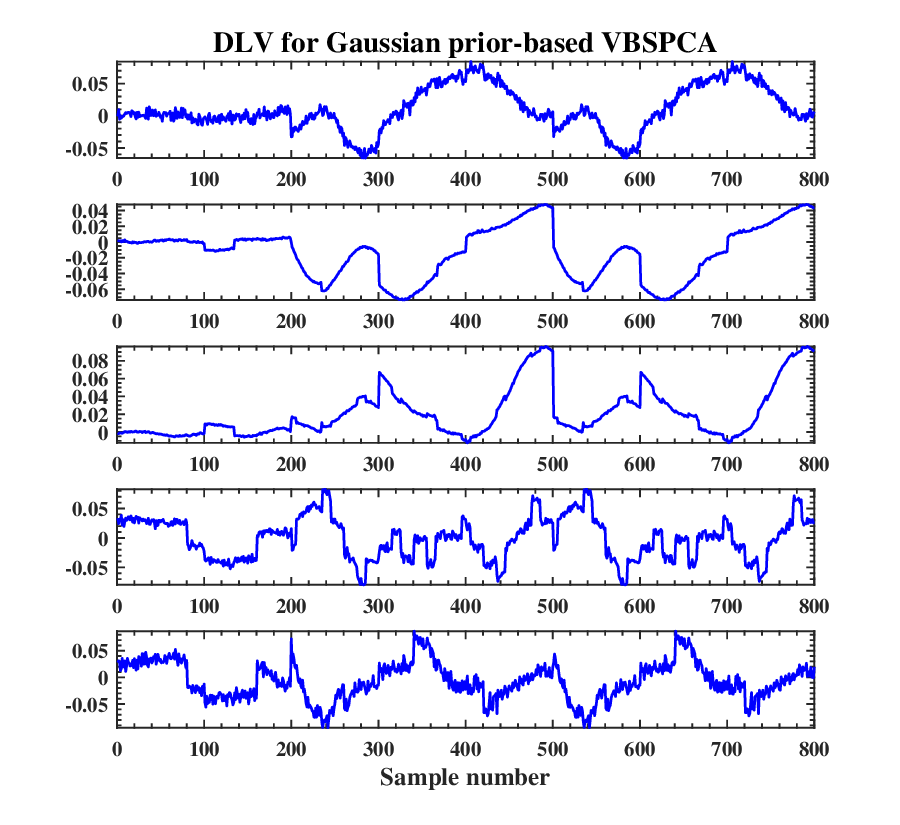}
	}
	\subfigure[Laplace prior-based VBSPCA.]{
		\includegraphics[width=3.3in]{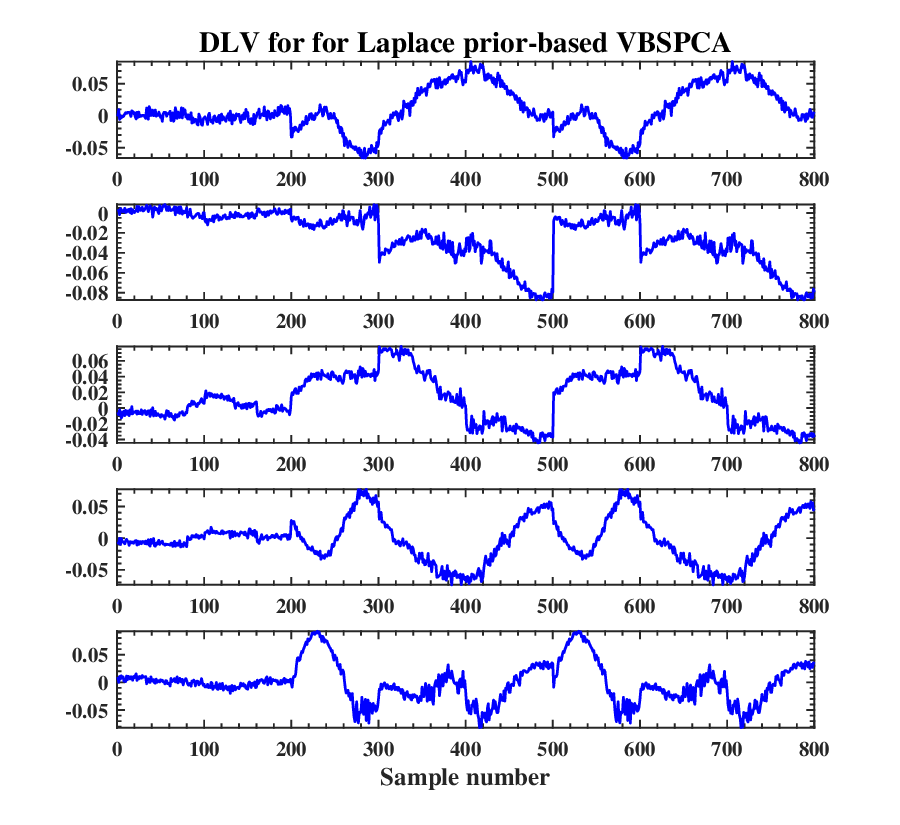}
	}
	\caption{Dynamic latent variables.}
	\label{fig.7}
\end{figure}

\begin{figure}[H]
	\centering
	\subfigure[Gaussian prior-based VBSPCA.]{
		\includegraphics[width=3.3in]{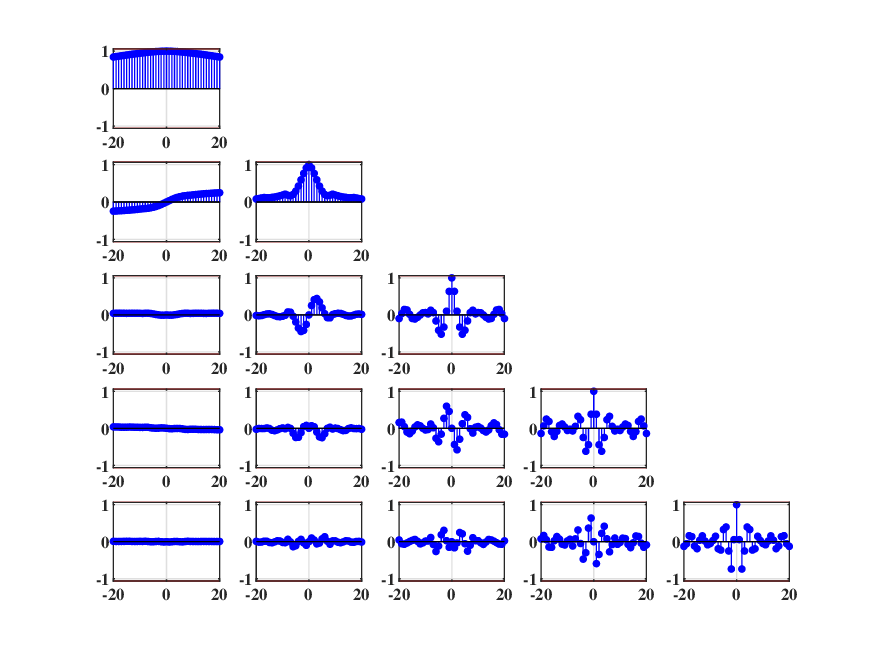}
	}
	\subfigure[Laplace prior-based VBSPCA.]{
		\includegraphics[width=3.3in]{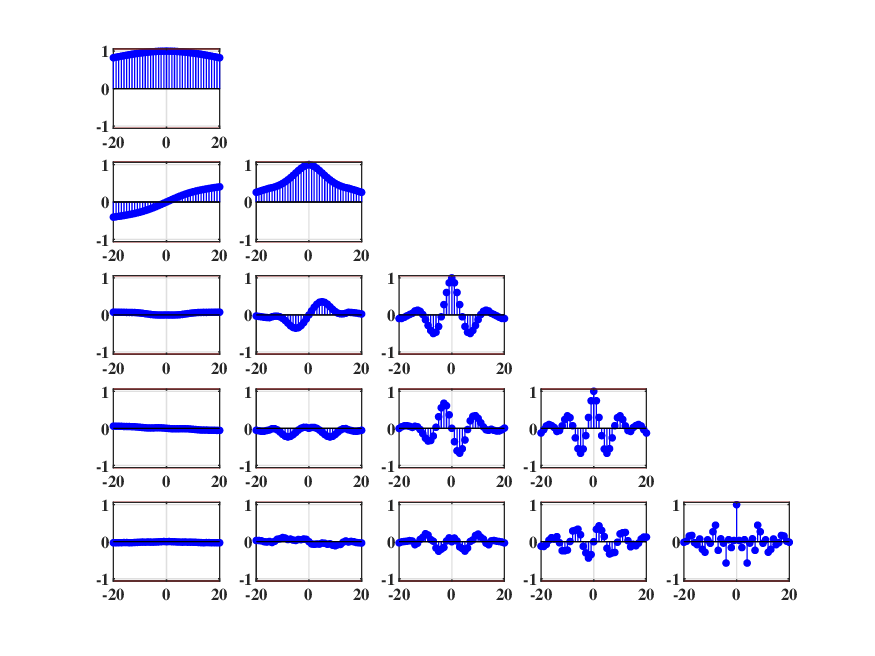}
	}
	\caption{Autocorrelation coefficients of dynamic latent variables.}
	\label{fig.8}
\end{figure}

	The predicted values of the latent variables calculated by the sparse VAR model are shown in Figure \ref{fig.7}, where the five dynamic latent variables have been able to characterize the full range of changes in the system. 
	It is also evident from the fault detection results that the first dynamic latent variable and the detection results exhibit the highest degree of similarity.
	The correlations of the obtained dynamic latent variables were checked by plotting correlation plots. The correlations among the dynamic latent variables are presented in Figure \ref{fig.8}. As depicted, it is evident that the correlations have been eliminated through modeling, whereby the first five dynamic latent variables are extracted.

\begin{figure}[H]
	\centering
	\subfigure[Gaussian prior-based VBSPCA.]{
		\includegraphics[width=3.3in]{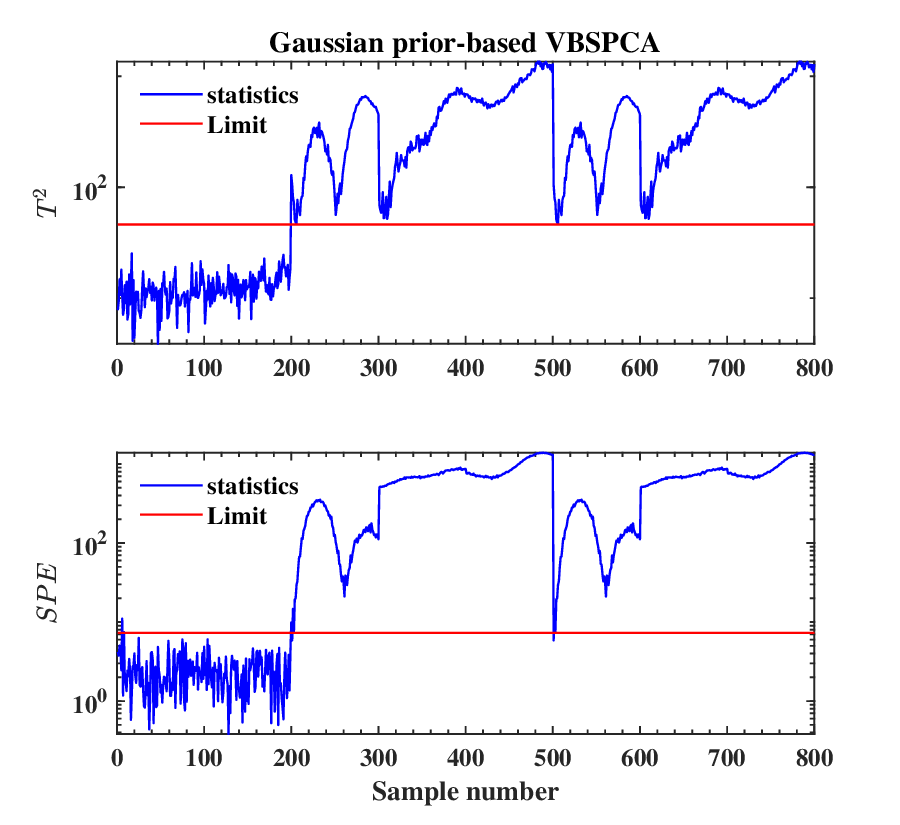}
	}
	\subfigure[Laplace prior-based VBSPCA.]{
		\includegraphics[width=3.3in]{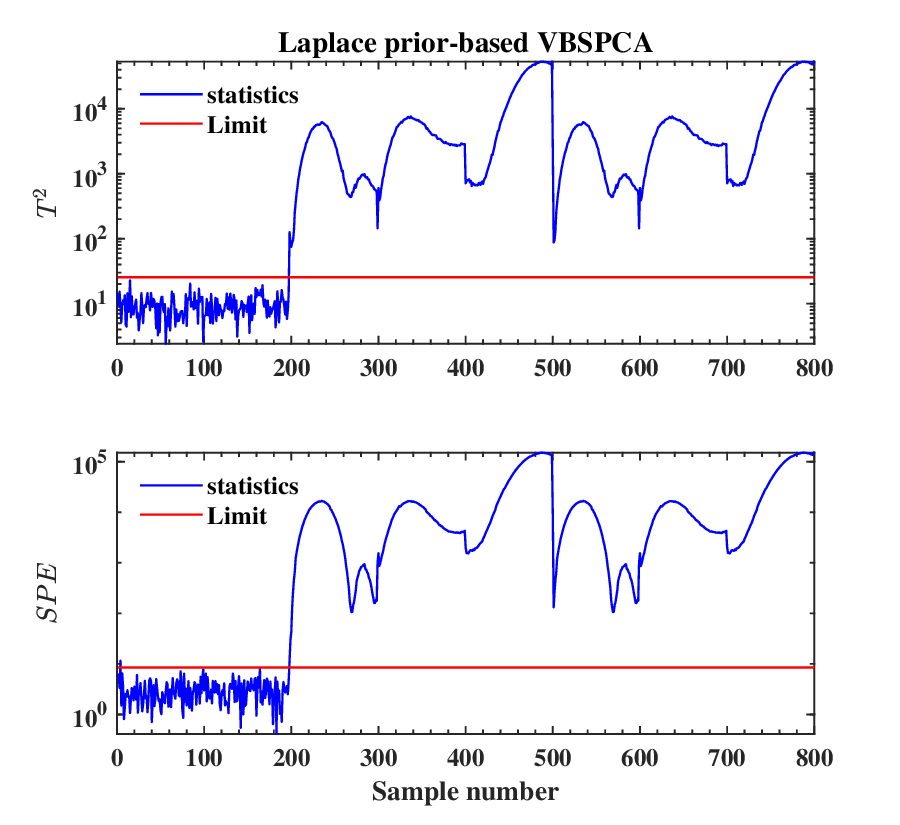}
	}
	\subfigure[Variational Bayesian dictionary learning.]{
		\includegraphics[width=3.3in]{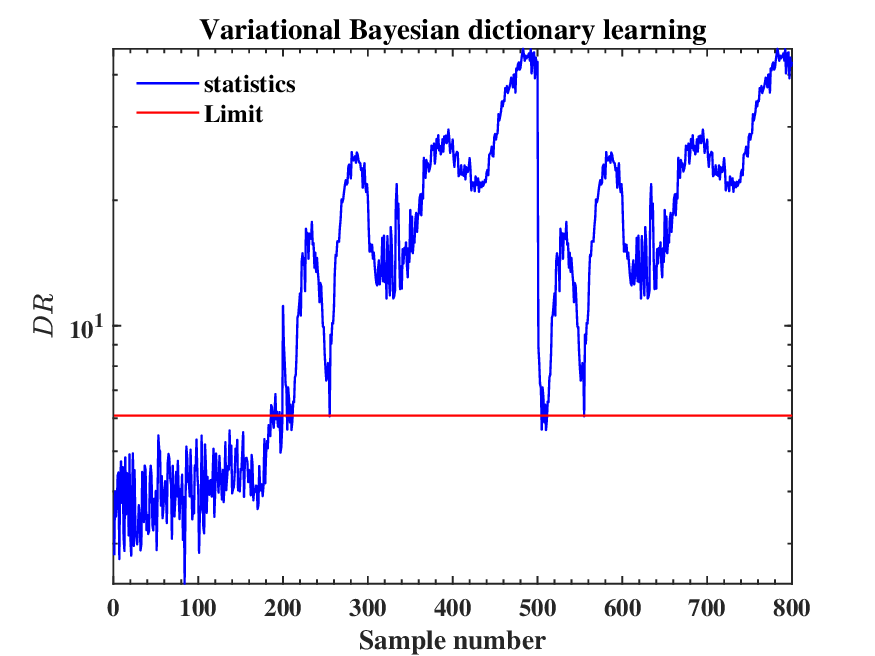}
	}
	\subfigure[Dynamic inner canonical correlation analysis.]{
		\includegraphics[width=3.3in]{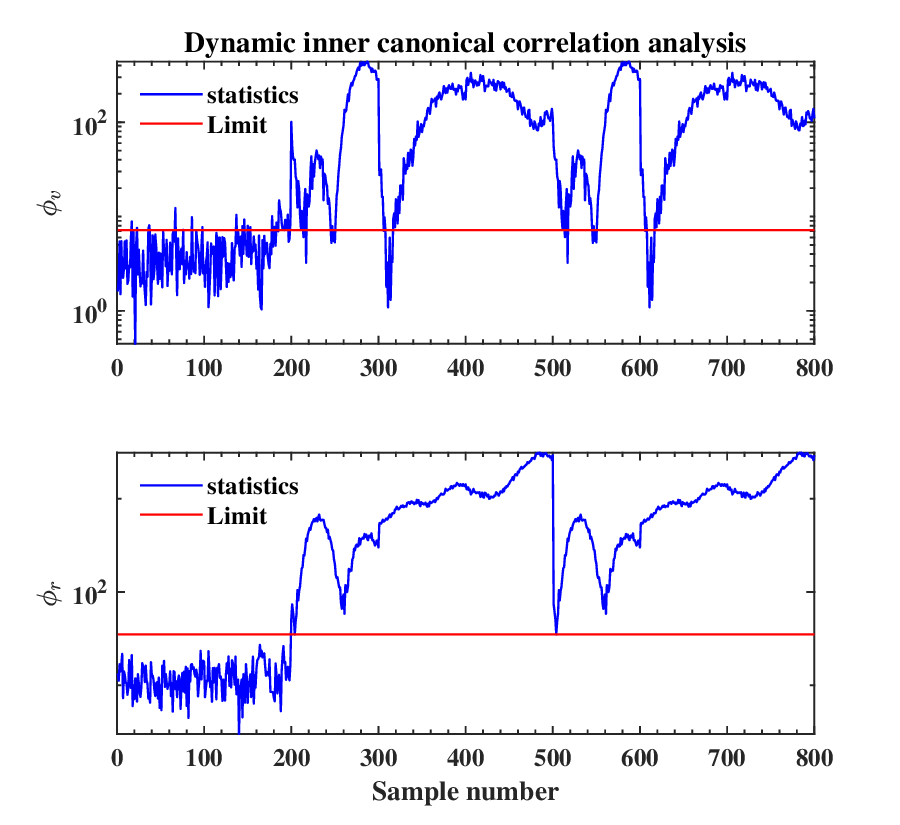}
	}
	\caption{Detection result of fault 6.}
	\label{fig.9}
\end{figure}

		The results of the fault detection are shown in Figure \ref{fig.9}.
	Fault 6 is also heavily contaminated by noise, and three different Bayesian sparsity methods effectively detect the fault, but the DiCCA method does not detect all faults. 
	VBDL fails to obtain sufficiently good detection results. This also shows the importance of dynamic analysis for fault detection.
	The Laplace prior-based VBSPCA method has a better fault detection ability compared to the  Gaussian prior-based VBSPCA method.
	This suggests that the Lasso sparsity corresponding to the Laplace prior is more favourable for feature selection and noise reduction.

\begin{figure}[H]
	\centering
	\subfigure[Gaussian prior-based VBSPCA with $T_2$.]{
		\includegraphics[width=3.3in]{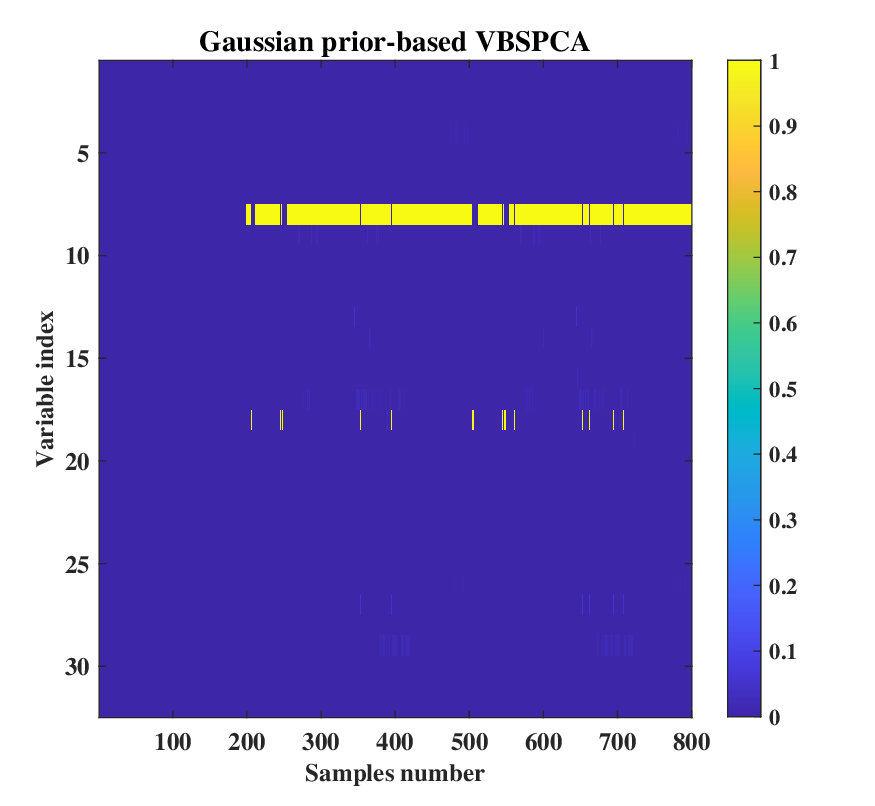}
	}
	\subfigure[Gaussian prior-based VBSPCA with $SPE$.]{
		\includegraphics[width=3.3in]{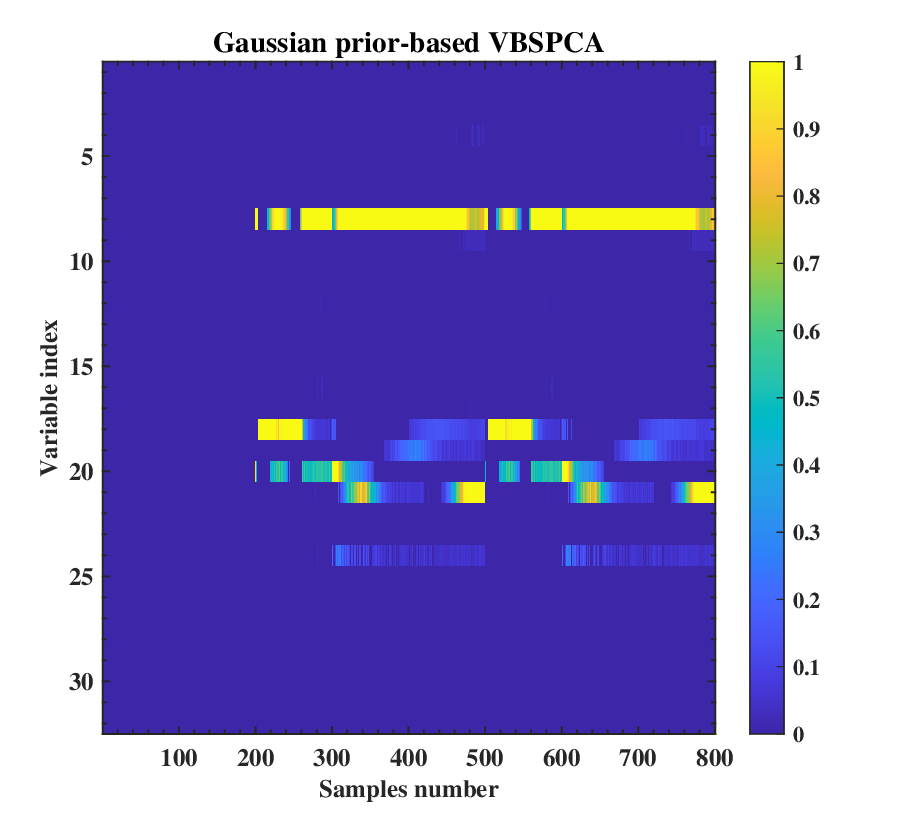}
	}
	\subfigure[Laplace prior-based VBSPCA with $T_2$.]{
		\includegraphics[width=3.3in]{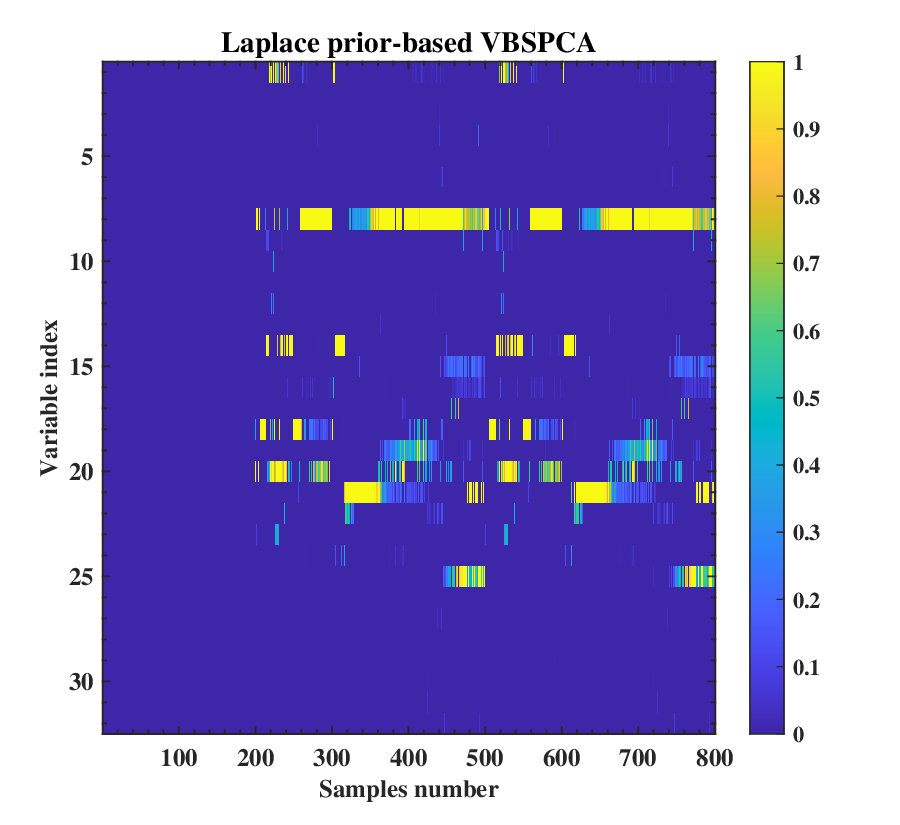}
	}
	\subfigure[Laplace prior-based VBSPCA with $SPE$.]{
		\includegraphics[width=3.3in]{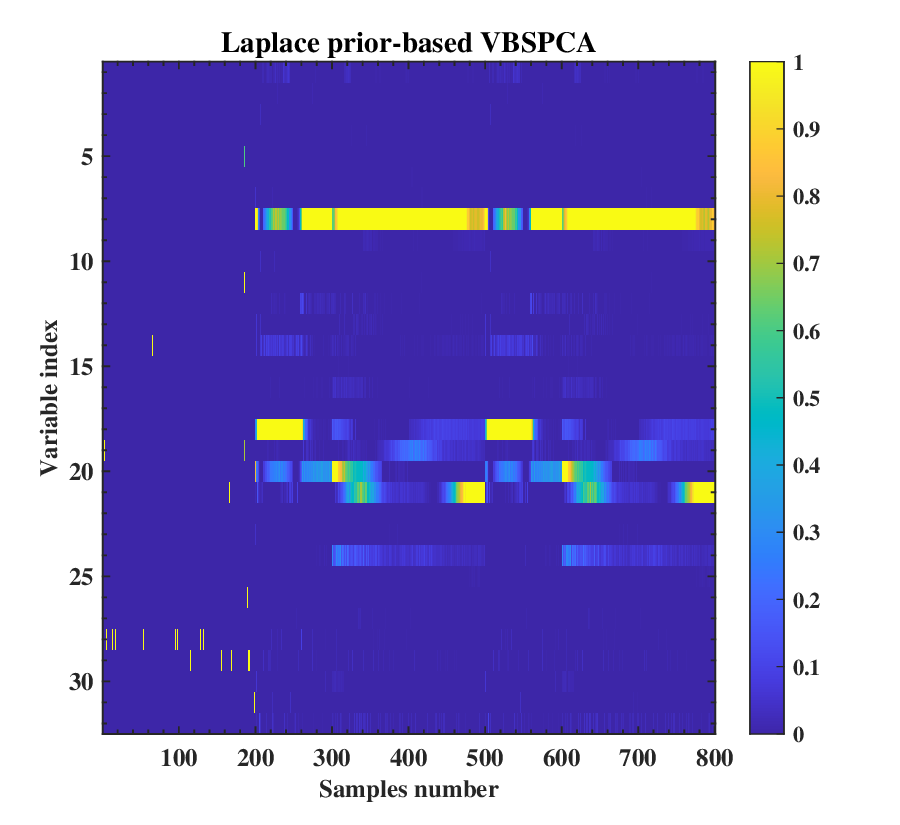}
	}
		\caption{Diagnosis result of fault 6.}
		\label{fig.10}
	\end{figure}

	The fault diagnosis results based on $T_2$ and $SPE$ indices are shown in Figures \ref{fig.10}.
	The fault diagnosis results indicate that in this case, fault 2 is primarily caused by variable 8, which makes an absolute contribution. It should be noted that under different conditions, this fault may be caused by different factors. Additionally, the contribution plot based on $SPE$ shows that variables 18 to 21 also contribute to the fault. 
	The Gaussian prior-based VBSPCA only considers variable 8 as contributing in $T_2$. But the Laplace prior-based VBSPCA assumes that there are many variables that contribute to the fault, although not continuously.

\section{Conclusions} \label{sec:conclusions}

Hydrogen production from electrolytic water is a green method of hydrogen production, which is the main way to produce hydrogen from renewable energy such as photovoltaic and wind power. 
With the rapid development of renewable energy and the update and iteration of material technology, electrolytic water hydrogen production technology has a broad development prospect. 
However, the safety of the hydrogen production process is often overlooked. 
The process monitoring of electrolytic water hydrogen production refers to the real-time measurement and control of each parameter in the electrolytic water hydrogen production system to ensure the safe, stable and efficient operation of the system.
The industrial environment causes the AWE process data to be randomly contaminated by noise.
In this study, VBSPCA approach is developed to detect fault of AWE industrial process.
We derive two sparse PCA methods in the framework of variational inference. 
And it is shown that Laplace prior and Gaussian prior correspond to $\ell_1$ regularisation and $\ell_2$ regularisation. 
Constructing a sparse projection matrix is more conducive to feature selection and improves the robustness of the model to noise.
The proposed method can detect faults in the production environment of industrial noise and has the capability to dynamically analyze data. 
The established dynamic process monitoring scheme can effectively extract the dynamic relationship of the data and obtain reliable detection and diagnosis results.
Therefore, the proposed solution is attractive for green hydrogen production applications.


\bibliographystyle{elsarticle-num}

\bibliography{VBSPCA}






\end{document}